\documentclass[10pt,journal,compsoc,twoside]{IEEEtran}
\ifCLASSOPTIONcompsoc
  \usepackage[nocompress]{cite}
\else
  \usepackage{cite}
\fi
\ifCLASSINFOpdf
\else
\fi
\usepackage{xcolor,graphicx}
\begin{document}

\title{Security of the Internet of Things:\\Vulnerabilities, Attacks and Countermeasures}

\author{Ismail~Butun,~\IEEEmembership{Member,~IEEE,}
        Patrik~\"{O}sterberg,~\IEEEmembership{Member,~IEEE,}
        \\and~Houbing~Song,~\IEEEmembership{Senior~Member,~IEEE}% <-this % stops a space
\IEEEcompsocitemizethanks{\IEEEcompsocthanksitem I. Butun is the corresponding author of this manuscript. \newline I. Butun is with Department of Computer Science and Engineering, Chalmers University of Technology, Gothenburg, Sweden.\protect\\
% 851 70 Sundsvall, SWEDEN.\protect\\	
E-mail: ismail.butun@chalmers.se
\IEEEcompsocthanksitem P. Osterberg is with Department of Information Systems and Technology (IST), Mid Sweden University, Sundsvall, Sweden.\protect\\
E-mail: patrik.osterberg@miun.se
\IEEEcompsocthanksitem H. Song is with Department of Electrical, Computer, Software, and Systems Engineering, Embry-Riddle Aeronautical University, FL, USA.\protect\\
% University, Daytona Beach, FL.\protect\\
E-mail: h.song@ieee.org
}% <-this % stops an unwanted space
\thanks{Manuscript received March 5, 2019.}}%; revised August 26, 2015.}}


\markboth{ to be appear at: IEEE Communications Surveys \& Tutorials,~Vol.~XX, No.~X, XX~2019}
{Butun \MakeLowercase{\textit{et al.}}: Security of the Internet of Things - DRAFT VERSION}




\IEEEtitleabstractindextext{%
\begin{abstract}
Wireless Sensor Networks (WSNs) constitute one of the most promising third-millennium technologies and have wide range of applications in our surrounding environment. The reason behind the vast adoption of WSNs in various applications is that they have tremendously appealing features, e.g., low production cost, low installation cost, unattended network operation, autonomous and longtime operation. WSNs have started to merge with the Internet of Things (IoT) through the introduction of Internet  access capability in sensor nodes and sensing ability in Internet-connected devices.  Thereby, the IoT is providing access to huge amount of data, collected by the WSNs, over the Internet. However, owing to the absence of a physical line-of-defense, i.e. there is no dedicated infrastructure such as gateways to watch and observe the flowing information in the network, security of WSNs along with IoT is of a big concern to the scientific community. More specifically, for the application areas in which CIA (confidentiality, integrity, availability) has prime importance, WSNs and emerging IoT technology might constitute an open avenue for the attackers. Besides, recent integration and collaboration of WSNs with IoT will open new challenges and problems in terms of security. Hence, this would be a nightmare for the individuals using these systems as well as the security administrators who are managing those networks. Therefore, a detailed review of security attacks towards WSNs and IoT, along with the techniques for prevention, detection, and mitigation of those attacks are provided in this paper. In this text, attacks are categorized and treated into mainly two parts, most or all types of attacks towards WSNs and IoT are investigated under that umbrella: ``Passive Attacks'' and ``Active Attacks''.  Understanding these attacks and their associated defense mechanisms will help paving a secure path towards the proliferation and public acceptance of IoT technology.
\end{abstract}

}

\maketitle

\IEEEdisplaynontitleabstractindextext

% \fi
\IEEEpeerreviewmaketitle

\IEEEraisesectionheading{\section{Introduction}\label{sec:introduction}}

%\IEEEPARstart{T}{his} demo file is intended to serve as a ``starter file''
%for IEEE Computer Society journal papers produced under \LaTeX\ using
%IEEEtran.cls version 1.8b and later.
%I wish you the best of success.

\IEEEPARstart{R}{ecent}  developments in wireless communications and Micro Electro Mechanical Systems (MEMS) technologies facilitated the design of Wireless Sensor Networks (WSNs), in which sensor nodes collect the intelligible data from their surrounding environments and share them in a wireless fashion to send the information towards a meaningful data sink. According to scientific predictions, the total number of wireless sensors deployed is expected to reach 60 trillion at the end of the year 2022, meaning 10 thousand sensors for every person on the world \cite{butun2013prevention}. Therefore, all the problems and challenges concerning WSNs will expose plentiful topics for the researchers.

\begin{figure*}[!t]
\centering
\includegraphics[width=\textwidth]{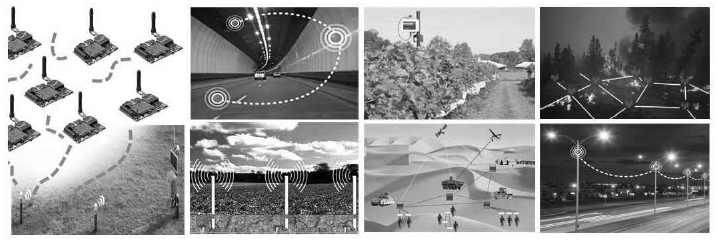}
\caption{Various application fields of IoT enabled WSNs.}
\label{fig1}
\end{figure*}

Owing to their easy and cheap deployment features, WSNs has wide-scale application areas in science as shown in Fig.~\ref{fig1}: To monitor environment-related events (such as wildfire, earthquake, ocean, pollution, water quality, wildlife), to collect information regarding human-related activities and observation of human behavior (such as elder-care, nursery, healthcare), to provide mission-critical information (such as military operations, highway traffic); to monitor industrial sites (such as industrial automation, manufacturing machinery performance), and so on \cite{butun2014survey}.

%\Figure[t!](topskip=0pt, botskip=0pt, midskip=0pt)[width=\textwidth]{figure1.png}
%{Various application fields of IoT enabled WSNs.\label{fig1}}

Internet of Things (IoT) is revolutionizing the IT sector and will be next big leap of the technology following Internet. IoT market is expected to grow from more than 15 billion devices in 2015 to more than 75 billion in 2025 \cite{dataeconomy}. This prediction means that on average, each person on earth will have at least 25 personal IoT devices in 7 years time. Henceforth, IoT is expected to have a dramatic impact on our lives in near future \cite{lin2017survey}. During this period, WSNs will be integrated into IoT and innumerable sensor nodes will join the Internet aiming at cooperating with other nodes to sense and monitor their environment. IoT will provide an interaction between people and environment by using the WSNs more and more in near future \cite{kocakulak2017overview}. For instance, our earth will benefit from this integration by the result of the increased environmental awareness \cite{fang2014integrated}.

The vision behind IoT is to let people and smart things to be connected at any time, in any place, to anything and anyone, via any network and service \cite{abomhara2014security}. So by following this vision, application areas of IoT will increase continuously and dramatically for every aspect of life. For example, nowadays, with the diverse installation of IoT devices, we are able to remotely sense and act upon situations regarding our houses or offices. E.g. in an event of intrusion to the premises, an alert can reach to our smartphone asking immediate attention or trigger an automatic response on our behalf. The pictures and videos being taken can be sent directly to the police so that they may approach the crime scene better prepared with the intelligence they gathered in advance.

Security in WSNs and IoT, is an important issue, especially if they are commissioned for mission-critical tasks. For instance, in tactical military applications where a security gap in the network could cause casualties for the friendly forces on a battlefield. Another example would be from health-care sector (IoT applications): A recent paper \cite{gope2016bsn} revealed that most of the current used systems fail to embed strong security services that could be preserve patient privacy. None of the patients would be happy if their confidential health data were exposed through the leakages from misbehaving nodes or due to system failures.

Algorithms and methodologies designed for securing WSNs will be relevant to any IoT that comprises one or more sensor networks. As also mentioned earlier, WSNs most probably will be integrated with IoT in the near future \cite{li2015internet}. Therefore, all cyber-security related issues, especially attacks, their prevention and mitigation are very important for establishing secure and reliable IoT. 

WSNs are vulnerable to a wide range of attack types which might put critical threats to the security of those networks. Therefore these attack types need to be investigated thoroughly. Security related attacks against WSNs can be branched into two main categories: Active attacks and Passive attacks. In the \textit{passive attacks} category, attackers are generally hidden (camouflaged) and either damage the functioning components of the network; or tap the communications link to collect useful information. Passive attacks can also be further categorized into eavesdropping, node destruction, node malfunctioning, node outage and traffic analysis types. Whereas in the \textit{active attacks} category, an attacker effects the functions and operations of the targeted network. The result of this ill-effect can be the real objective of the attacker and can also be detected by security mechanisms (intrusion detection). For instance, network services might be vulgarized as a consequence of this type of attacks. Active attacks can also be further categorized into jamming, flooding, Denial-of-Service (DoS), blackhole, wormhole, sinkhole and Sybil types.

In Computer Science, solutions to defend against security attacks towards networks comprise of three essential components \cite{butun2014survey}:
\begin{itemize}
	\item Prevention: This component aims at preventing attacks before they happen. In this case, any proposed method needs to be able to devise measures to defend against the specific type of attack(s). Intrusion prevention mechanisms can resist external attackers towards WSNs and IoT, but they are not specifically designed to resist the internal attackers.
	\item Detection: In an event of an attack, if an adversary manages to advance the measures taken by the prevention component, this means that defense against the attack has failed. At the moment, security solutions that are devised for the detection component of the related attack would take in-charge and work at especially in identifying those nodes that are compromised. The only way of reacting against ongoing attacks, especially internal attacks, is using the Intrusion Detection Systems (IDSs). After an intrusion is detected, then a mitigation mechanism would be issued to minimize the adverse effects of the ongoing attack. 
	\item Mitigation: Final component aims at mitigating attacks after they happen, for example, in order to secure network, a security measure should be taken, such as `dismissing the affected nodes in a network' or `disabling the ports of a computer which were used during the attack'. 
\end{itemize}
Thereby, all these three components constitute a whole security structure and cannot be considered separately in defending WSNs and IoT against diverse kinds of attacks.

In the literature, various surveys are provided to present security issues in WSNs: Butun \textit{et al.} \cite{butun2014survey} provided a survey of intrusion detection systems, Zhu \textit{et al.} \cite{zhu2012detecting} provided a survey of detecting node replication attacks, Chen and Chao \cite{chen2014survey} provided a survey of key distribution, Han \textit{et al.} \cite{han2014management} provided a survey of trust management. Finogeev \textit{et al.} \cite{finogeev2017information} provided a survey on attacks and security in WSNs of industrial SCADA systems. Following publications provided limited surveys of security issues and attacks against WSNs (and IoT) and their classifications: Padmavathi and Shanmugapriya \cite{padmavathi2009survey}, Pathan \textit{et al.} \cite{pathan2006security}, Shabana \textit{et al.} \cite{shabana2016security}, Bartariya and Rastogi \cite{bartariya2016security}, Sharma and Ghose \cite{sharma2010wireless}, and Borgohain \textit{et al.} \cite{borgohain2015survey}. However, according to the best of our knowledge, the survey provided in this paper is the most comprehensive and the detailed one covering all the attacks towards WSNs along with their related detection, prevention, and mitigation techniques. Besides, our paper also provides a path to defend IoT, by considering the lessons learned while securing WSNs. 

Security analysis of sub-domains of IoT, such as LPWAN networks, is omitted in this manuscript. E.g., readers that are interested in the security of LoRaWAN can refer to following works: Butun \textit{et al.} \cite{butun2018analysis,butun2019security}, Eldefrawy \textit{et al.} \cite{eldefrawy2018formal}, Haxhibeqiri \textit{et al.} \cite{haxhibeqiri2018survey}, and  Sinha \textit{et al.} \cite{sinha2017survey}. Privacy and trust related issues of IoT are also omitted in this text, in order to keep the focus on ``attacks and mitigation''. Readers that are interested in that specific topic may refer to following works: Butun \cite{butun2017privacy}, Chen \textit{et al.} \cite{chen2016trust}, Ott \textit{et al.} \cite{ott2015trust}, Sicari \textit{et al.} \cite{sicari2015security}, and Yan \textit{et al.} \cite{yan2014survey}.

Security of the IoT is a very wide (attacks and their counter-measures, privacy, trust, key-distribution, patch-management, access-control, etc.) and also an emerging topic. Hence, the aim of this survey is to present all cyber-security attacks against WSNs and IoT along with their related defense mechanisms. We believe that this would shed light on researchers who are considering to devise security algorithms for IoT. For this sake, we also provide a section discussing the state-of-art networking technologies in IoT. However, additional reading is advised as follows: Shelby \textit{et al.} \cite{shelby2014constrained} and Hartke \cite{hartke2015observing} for CoAP, Banks \textit{et al.} \cite{banks2014mqtt} and Yokotani \textit{et al.} \cite{yokotani2016comparison} for MQTT, Brandt \textit{et al.} \cite{brandt2016applicability} and Zhang \textit{et al.} \cite{zhang2014evaluating} for RPL, Nikshepa \textit{et al.} \cite{nikshepa20186lowpan} and Fabre \textit{et al.} \cite{fabre2016deploying} for 6LoWPAN, Dujovne \textit{et al.} \cite{dujovne20146tisch} and Watteyne \textit{et al.} \cite{watteyne2017teaching} for 6TiSCH, and finally Chang \textit{et al.} \cite{chang2015adaptive} and Watteyne \textit{et al.} \cite{watteyne2015using} for TSCH.

In this survey, prevention, detection, and mitigation of attacks towards WSNs and IoT is the topic of interest. Therefore, the rest of the paper is organized as follows: Section~\ref{IoT} briefly overviews the definition of IoT along with trends, impact and future projections. Section~\ref{sec:attacks} presents various types of attacks towards the WSNs and IoT. Section~\ref{sec:defence} provides the defense strategies including prevention, detection, and mitigation, against those attacks mentioned in Section~\ref{sec:attacks}. Cyber-security of IoT including open challenges, cyber-attacks, and defense mechanisms are discussed in Section~\ref{sec:iot}. Section~\ref{sec:discussion} presents unique security solutions in the field, discusses the inclusion of security during WSN-IoT integration, and presents final remarks. Section~\ref{sec:conclusion} concludes the paper. List of abbreviations is presented in the Appendix section.

\subsection{Internet of Things: Definition, Trends, Impact, and Future Projections}\label{IoT}

%\subsubsection{Industry 4.0 and IIoT}
%
%
%\subsubsection{IIoT}
%
%\subsubsection{Security Threats and Vulnerabilities}
%Botnets and Malware

\begin{figure}[!t]
	\centering
	\includegraphics[width=2.5in]{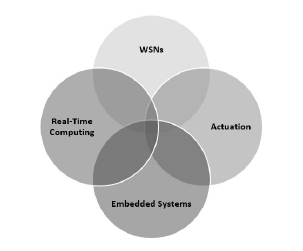}
	\caption{IoT is the Confluence of Several Technologies.}
	\label{fig2}
\end{figure}

The term "Internet of things" was coined by Kevin Ashton of Procter \& Gamble, later MIT's Auto-ID Center, in 1999. Since then, the Internet of Things (IoT) has rapidly evolved into a field that involves the interconnection and interaction of smart objects, which are objects or devices with embedded sensors, on-board data processing capability, and a means of communication, to provide automated services and applications \cite{jeschke2017industrial,lin2017survey, forsstrom2018challenges}. Rather than a single technology, IoT involves the convergence of WSNs, real-time computing, embedded systems, and actuation technologies \cite{song2016cyber}, as shown in Fig.~\ref{fig2}.

Today, most of what we consider as IoT is a variety of largely stand-alone devices and isolated systems, such as wearable fitness monitors \cite{zhang2014ubiquitous}, smart watches, smart phones, home thermostats and lighting \cite{xu2016toward}, and remote video streaming \cite{rani2017iomt}. Emerging IoT implementations will use smaller and more energy- efficient embedded sensor technologies, enhanced communications, advanced data analytics, and more sophisticated actuators to collect and aggregate information and enable intelligent systems that understand context, track and manage complex interactions, and anticipate requirements \cite{sun2016internet, hu2017security, xu2017security, zhu2018new,butun2018security}. 

IoT is expected to become ubiquitous, with implementations in the smart home for management of energy use, control of appliances, monitoring of food and other consumables \cite{xu2016toward, rani2017iomt, song2017security, song2017smart}; consumer applications such as health and fitness monitoring, condition diagnosis \cite{zhang2014ubiquitous}; manufacturing and industrial settings for supply chain management, robotic manufacturing, quality control, health and safety compliance \cite{jeschke2017industrial}; utility grids and other critical infrastructure for grid optimization, automated fault diagnosis, automated cyber security monitoring and response \cite{song2017security}; and automotive/transportation for optimization for driving conditions, assessing driver alertness, collision/accident avoidance, and managing vehicle health \cite{sun2017secure, li2018policy}.

IoT is a networking infrastructure for Cyber-physical systems (CPS) \cite{jeschke2017industrial}, which are engineered systems that are built from, and depend upon, the seamless integration of computation and physical components \cite{song2016cyber}. Advances in CPS will enable capability, adaptability, scalability, resiliency, safety, security, and usability. CPS technologies are transforming the way people interact with engineered systems, just as the Internet has transformed the way people interact with information. CPS have been applied successfully in a range of application domains including agriculture, aeronautics, building design, civil infrastructure, energy, environmental quality, healthcare and personalized medicine, manufacturing, and transportation \cite{song2016cyber}. 

\begin{figure*}[!t]
	\centering
	\includegraphics[width=\textwidth]{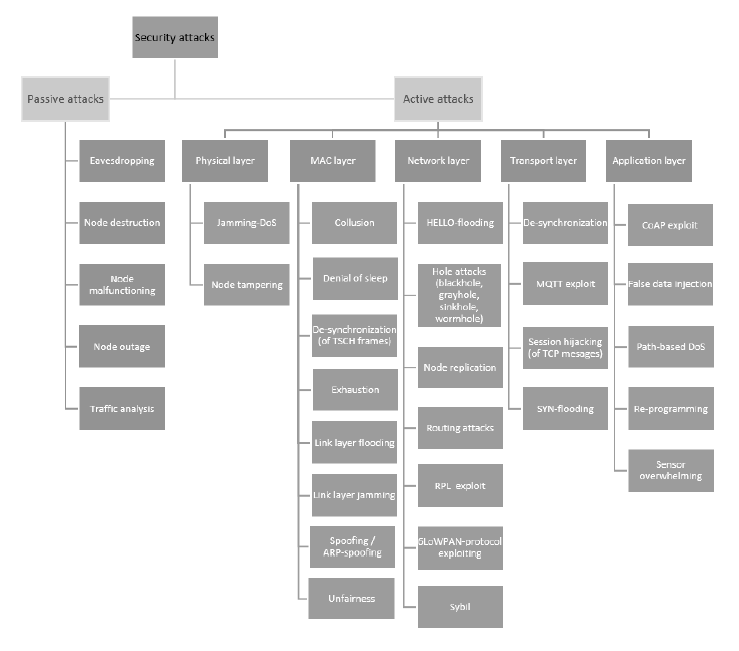}
	\caption{Security attacks towards the WSNs and IoT - OSI stack protocol layered description.}
	\label{fig3}
\end{figure*}

Innovations in IoT potentially impact a variety of applications and services, such as connected cities and homes, smart transportation, smart agriculture, industrial IoT, and retail IoT. IoT enabled smart and connected communities will bring about new levels of economic opportunity and growth, safety and security, health and wellness, and overall quality of life \cite{song2017smart}. IoT enabled smart transportation will provide improved safety, mobility, and energy conservation in the development and operation of the highway system \cite{sun2017secure}. Smart agriculture will deliver food, fiber, fuel, and feed within a changing global climate while reducing agriculture's environmental footprint and managing biotic threats to production \cite{song2016cyber}. Industrial IoT will enable an increasingly wide range of value-added manufacturing services by being intelligent, precise, predictable, reliable, secure, and adept with fabricating new materials; connected and broadly accessible, with capabilities that are transparent to users; connected to applications that reside in the cloud and plug into an expansible, interactive architecture; accessible at low cost to innovators and entrepreneurs, including both users and providers of manufacturing services; clean, green, and resource-efficient; and resilient to disruptions \cite{jeschke2017industrial}. Retail IoT will increase business efficiency, drive more sales and improve customer satisfaction \cite{song2017smart}. 

%%\hfill mds
% 
%%\hfill August 26, 2015
%
%\subsection{Subsection Heading Here}
%Subsection text here.
%
%% needed in second column of first page if using \IEEEpubid
%%\IEEEpubidadjcol
%
%\subsubsection{Subsubsection Heading Here}
%Subsubsection text here.

\section{Attacks towards the WSNs and IoT}\label{sec:attacks}
In the literature, there is a variety of classifications for attacks towards the WSNs \cite{padmavathi2009survey,wood2002denial,huang2003cooperative,karlof2003secure,raymond2008denial,ccayirci2009security,hu2005security,znaidi2008ontology}. Among these, we will consider the activity of the attacker (passive/active) as main categorization and the targeted Open Systems Interconnection (OSI) model (layered description of stack protocol) as sub-classifications as shown in Fig.~\ref{fig3}. Following subsections include descriptions of each item in the Fig.~\ref{fig3}:

\subsection{Passive Attacks}
Passive attacks are performed in a way that it cannot be sensed by any means. This is because of the fact that the adversaries do not make any radio emissions. Since wireless links are easier to tap, wireless networks are more susceptible to passive attacks, such as eavesdropping, which can be performed easily listening to the wireless communication amongst sensor nodes in the WSN without capturing any of them. Passive attacks are mainly against data confidentiality.

In passive attacks, attackers are typically camouflaged, i.e. hidden, and tap the communication lines to collect data. Passive attacks can be grouped into eavesdropping, node malfunctioning, node tampering/destruction, node outage and traffic analysis types (see Fig.~\ref{fig3}) \cite{butun2013prevention}.

Here, it is important to state that node malfunctioning, node outage and node tampering are considered as \textit{active attacks} in some papers \cite{padmavathi2009survey,wood2002denial,huang2003cooperative,karlof2003secure,raymond2008denial,ccayirci2009security,hu2005security,znaidi2008ontology}. We present them as \textit{passive attacks} in this paper, hence they do not introduce big concern (they do not constitute a single point of failure, as the network can continue its operation without the contribution of the failed nodes!) to the network compared to other more impactive active attacks. 

\subsubsection{Passive Information Gathering (Eavesdropping)}
Eavesdropping is also known as ``Passive information gathering''. Classified data can be eavesdropped by tapping communication lines. Hence wireless links are easier to tap, wireless networks are more susceptible to passive attacks. Since WSNs use short-range communications, an attacker must be in proximity in order to gather useful information by eavesdropping. WSNs are a little more secure against tapping compared to other long-range wireless technologies because signals are sent over shorter distances. Interception of the messages transmitted through WSNs might reveal following useful information: physical location of specific nodes such as cluster heads, gateways, key distribution centers, etc.; message identities (IDs), timestamps, other fields, almost anything that is not encrypted.

\subsubsection{Node Destruction}
Physical destruction (with the usage of electrical surge, physical force or ammunition) of the nodes by any means.

\subsubsection{Node Malfunctioning}
This may happen due to many different factors from faulty sensors or energy depletion due to sensor overwhelming or other DoS attacks. 

\subsubsection{Node Outage}
This attack occurs whenever a node fails its regular functionality. For example, if a cluster head of a heterogeneous network fails at regular operation, then the WSN protocols have to be strong enough to mitigate the negative effects of this kind of node outages, by electing new cluster heads and/or providing alternate routes for network paths.

\subsubsection{Traffic Analysis}
The traffic pattern of a network may be as valuable as the content of data packets for adversaries. Important information about the networking topology can be derived by analyzing traffic patterns. In WSNs, the nodes closer to the base station, i.e. the sink, make more transmissions than the other nodes because they relay more packets than the nodes farther from the base station. Similarly, clustering is an important tool for scalability in WSNs and cluster heads are busier than the other nodes in the network. Detection of the base station, the nodes close to it or cluster heads may be very useful for adversaries because a denial-of-service attack against these nodes or eavesdropping the packets destined for them may have a greater impact. By analyzing the traffic, this kind of valuable information can be derived. Moreover, traffic patterns can pertain to other confidential information such as actions and intentions. In tactical communications, silence may indicate preparation for an attack, a tactical move or infiltration. Similarly, a sudden increase in the traffic rate may indicate the start of a deliberate attack or raid.

\subsection{Active Attacks}
In the active attacks, malicious acts are carried out not only against data confidentiality but also data integrity. Active attacks can also aim for unauthorized access and usage of the resources or the disturbance of an opponent's communications. An active attacker makes a radio emission or action that can be sensed by the WSN elements \cite{ccayirci2009security}. An example is DoS attack in the physical and/or network layer that would cause network elements to drop data packets. 

A Denial-of-Service (DoS) attack mainly targets the availability of network services. A DoS is generally explained as any kind of situation that consumes resources and diminishes the capacity of a network, therefore diverts the network from performing its expected functionality correctly or in a timely manner. A node is isolated from the rest of the network by blocking the incoming and outgoing packets.  In DoS attack, an adversary attempts to prevent legitimate and authorized users of services offered by the network from accessing those services. The classic way to achieve this is to flood packets to any centralized resource (access point) used in the network so that the resource is no longer available to the nodes in the network, resulting in the network no longer operating what was designed for. This may lead to a failure in the delivery of guaranteed services to the end users. 

In the active attacks, an adversary actually affects the operations of the attacked network. This effect may be the objective of the attack and can be detected. For example, the networking services may be degraded or terminated as a result of these attacks. Sometimes the adversary tries to stay undetected, aiming to gain unauthorized access to the system resources or threatening confidentiality and/or integrity of the content of the network. Active attacks of our interest (for WSNs) are grouped into five main groups, by following the OSI stack protocol layered structure, as shown in Fig.~\ref{fig3}. 

The OSI network structure consists of 5 layers for WSNs (and IoT) as described in \cite{kocakulak2017overview}: Physical, Data-Link (MAC), Network, Transport, and Application. It should be noted that Session and Presentation layers of the traditional OSI network model are all considered in the Application layer of WSNs (and IoT).

\subsubsection{Attacks towards Physical Layer}
\paragraph {Jamming DoS}
It is a DoS attack at the physical layer \cite{hu2005security}. A malicious device can jam a signal by transmitting at the same frequency. The jamming signal contributes to the noise in the carrier and its strength is enough to reduce the signal-to-noise ratio below the level that the nodes using that channel need to receive data correctly. Jamming can be conducted continuously in a region, which thwarts all the nodes in that region from communication. Alternatively, jamming can be done temporarily with random time intervals, which can still very effectively hamper the transmissions.

\paragraph{Node Capture (Tampering)}
An adversary takes over the control of the sensor node by a physical attack, e.g. attaching cables to its circuit board and reading stored data as well as ongoing transmission in the WSN \cite{butun2013prevention}. Besides, by tampering adversaries can change the original wiring of the electronic board or change the content of the memory of the nodes and use the captured slave node by any means. Capturing a node might expose its critical data, especially revealing of cryptography-related keys and therefore might cause compromisation of the whole WSN. Two problems arise in this case:
\begin{itemize}
	\item Captured node can make arbitrary queries on behalf of the attacker (DoS attack against availability). 
	\item Captured node can provide false data to the legitimate users (attack against integrity).
\end{itemize}

\subsubsection{Attacks towards Data Link Layer}
The algorithms in the data link layer, especially MAC schemes, present many exploitation opportunities for DoS attacks. For example, MAC layer DoS attacks may continuously jam a channel. More complex DoS attacks can be designed based on MAC layer addressing schemes. Data link layer attacks are categorized as follows: Collision, denial of sleep, de-synchronization, exhaustion, flooding, link layer jamming, spoofing, and unfairness.

\paragraph{Collision}
In collision attack, an adversary starts transmitting packets from the same channel of a legitimate node of the network, whenever the legitimate one starts transmission. Hence, as a result, both transmitted packets collide and the targeted receiver does not receive the whole meaningful packet from the transmitter due to the collision loss in the transmission. Hence it is useless, the received packet is discarded and the transmitter is asked for re-transmission of the packet \cite{borgohain2015survey}. Causing collusions of a single byte of a message would be sufficient to cause a CRC (Cyclic Redundancy Check) error and eventually corrupt the whole message. From the attacker point of view, the collision attack is more advantageous compared to the jamming attack, since consumed transmission energy is lower (because the radio is used just only for a short duration of time) as well as the probability of detection \cite{znaidi2008ontology}.

\paragraph{Denial of Sleep (Sleep Deprivation Torture)}
Preventing a node from going to sleep leading to energy depletion from draining the battery. This can be from collision attacks or repeated handshaking i.e. Request to Send (RTS)  and Clear to Send (CTS) flow control signals. In this attack, a node is forced to deplete whole energy stored in its batteries \cite{stajano2002resurrecting}.

\paragraph{De-synchronization}
Time Synchronized Channel Hopping (TSCH) is a MAC layer protocol presented in IEEE 802.15.4e standard. It empowers extreme consistency and possesses small duty cycles through the time synchronization and channel hopping techniques.\cite{sajjad2014security}
Attacks against the TSCH time synchronization can happen when an attacker transmits the messages in the time-slots that are alloted to the other users. This causes the packets to collide and to be lost. After carefully observing the back-off times an attacker can cause a series of these events which eventually would cause the neighboring motes to be  de-synchronized. Hence, this attack can be taught of as an advanced version of collusion attack.

\paragraph{Exhaustion}
If the collusion attack described above, continues until the targeted node depletes its energy, this is called \textit{exhaustion attack} \cite{znaidi2008ontology}. This kind of attack can be executed by using an ordinary node or a laptop, which have the ability to transmit radio signals in the same band as the rest of the sensors do.

\paragraph{Link Layer Flooding}
In this type of attack, a malicious node abuses the fairness of medium access by sending excessive MAC data packets or MAC control packets to its neighboring nodes. In the end, victim nodes suffer from DoS or the power of their batteries get exhausted. Additionally, this attack may also exhaust channel bandwidth resources.\cite{liu2005mac}

\paragraph{Link Layer Jamming}
In this type of attack, the most useful packets, i.e. data packets, are targeted to be jammed. The probability distribution of the packet arrival times is acquired and used against the packets transmission. This attack is shown to be successful against these MAC protocols: B-MAC, L-MAC, and S-MAC \cite{znaidi2008ontology}.

\paragraph{Spoofing/ ARP-Spoofing}
In the spoofing attack, a malicious node spoofs MAC address of another victim node and then creates a number of various legitimate identities out of the victim node and uses these identities anywhere else in the network \cite{shabana2016security}. Whereas in ARP-spoofing attack, an attacker sends spoofed ARP (Address Resolution Protocol) messages into the network. Generally, the aim is to associate the attacker's MAC address with the IP address of a higher ranked node such as the default gateway, causing any traffic meant for that IP address to be sent to the attacker instead.

\paragraph{Unfairness}
Sporadic usage of \textit{exhaustion attack} or mis-usage of cooperative MAC protocols can cause unfairness in the network \cite{wood2002denial}. Unlike a DoS attack, this attack does not cause a user or a node completely to be disconnected from the network, but it causes intermittent blackouts in which users send/receive delayed messages. This attack degrades the quality of service in the network, hence it provides an advantage to the least number of the sensor nodes and disadvantage to the rest of the network, as the rest of the nodes miss their transmission deadlines in real-time MAC protocol configuration.

\subsubsection{Attacks towards Network Layer}
In the case of \textit{network layer attacks}, an attacker injects a significant amount of packets into the network which causes congestion in the network traffic as well as deprivation of power resources throughout the network. Example: ``Routing table overflow attack: Creation of the routes to the non-existing nodes'' \cite{yang2004security}. Network layer attacks are categorized as HELLO flooding, hole attacks (blackhole, sinkhole, wormhole), node replication, routing, selective forwarding, and Sybil types.

\paragraph{HELLO-flooding}
In this kind of attack, an attacker (has longer transmission range than normal nodes) broadcasts advertisement messages to the whole network and convinces other nodes that it is located in their neighborhood. 

In a more technical description; routing protocols broadcast ``HELLO'' message to inform of their presence to one-hop neighbors. A node receiving such a packet assumes that it is within the radio range of the sender which may not be true during this attack. A malicious node may flood ``HELLO'' packets with high enough transmission power to convince every node in the network that it is their neighbor. When the other nodes send their packets to the malicious node, those packets are not received by any node.

Many network and MAC layer protocols ask nodes to broadcast ``HELLO'' packet for announcing their presence towards their neighbors. Any node, receiving such a packet might consider that it is enclosed in the normal radio-range of the packet sender. This assumption would be falsified in some specific cases as follows: A laptop-class attacker broadcasting routing or other information with large enough transmission power could convince every node in the network that the adversary is its neighbor.

``Flooding'' is usually used to denote the epidemic-like propagation of a message to every node in the network over a multi-hop topology. In contrast, despite its name, the HELLO flood attack uses a single hop broadcast to transmit a message to a large number of receivers \cite{karlof2003secure}.

\paragraph{Hole Attacks}
\begin{itemize}
	\item \textbf{Blackhole:} A malicious node may drop all the packets that it receives for forwarding. This attack is especially effective when the blackhole node is also a sinkhole. Such an attack combination may stop all the data traffic around the blackhole. In some texts, this attack is also referred as ``Selfishness''.
	
	\item \textbf{Sinkhole:} A malicious node can advertise by broadcasting to all the neighbor nodes that it is the best next hop for sending the packets to its destination. When a node becomes a sinkhole, it becomes the hub for its vicinity and starts receiving all the packets going to the base station. All traffic of the network is directed to this single node but in this case, sinkhole node does not drop any packets. By this way, it expects to remain undetected by the IDS. This creates many opportunities for any follow-on attacks. Since all traffic of the network passes through this particular node which literally ``sinks'' all the data it receives, the name is given to this attack.
	
	\item \textbf{Selective Forwarding (Grayhole):} It is a special kind of blackhole attack, in which malicious node acts more cleverly and does not drop every packet it receives but the ones it selects. By this way, attacker expects to remain undetected by the IDS. This type of attack is also called ``grayhole attack'' as it is a variant of blackhole attack.
	
	\hspace{0.1cm} Similar to sinkhole attacks, a malicious node subverts the routing protocol by making itself part of many routes but instead of dropping of all packets selectively drop some packets while forwarding others in order to avoid detection. Forwarding packets is a major responsibility of a routing node. However, a malicious node intentionally may drop any packet and forward other ones.
	
	\hspace{0.1cm} Multi hopped networks are generally built upon the following assumption: The participant sensor nodes would be faithful in forwarding the messages they receive. In a selective-forwarding attack, adversary nodes might reject forwarding some certain messages by simply dropping them and making sure that these packets are not distributed anymore. As an example of this kind of attack, a malicious node behaves like a blackhole and refuses to forward every packet it receives. However, such an attacker has the following risk: Neighboring nodes will conclude that it has failed and they may decide to seek another route. A more subtle form of this attack is when an adversary selectively forwards packets. An adversary interested in suppressing or modifying packets originating from a select few nodes can reliably forward the remaining traffic and limit suspicion of his/her wrongdoing.
	
	\item \textbf{Wormhole:} A tunnel (out of the band fast transmission path) is created between two nodes that can be utilized to transmit packets in a faster way. This way, two far parts of the network advertised as neighbors to attract the surrounding traffic \cite{hu2003packet}.
	
	\hspace{0.1cm} A malicious node can eavesdrop or receive data packets at a point and transfer them to another malicious node, which is at another part of the network, through an out-of-band channel. The second malicious node then replays the packets. This makes all the nodes that can hear the transmissions by the second malicious node believe that the node that sent the packets to the first malicious node is their single-hop neighbor and they are receiving the packets directly from it. The packets that follow the normal route reach destination node, later than those conveyed through the wormhole and are therefore dropped because they do more hops - wormholes are typically established through faster channels. 
	
	\hspace{0.1cm} Wormholes are very difficult to detect and can impact on the performance of many network services such as time synchronization, localization, and data fusion.
\end{itemize}

\paragraph{Node-Replication (Clone)}
An attacker intentionally puts replicas of a compromised node in many places in the network to incur inconsistency. Node-replication (clone) attack is one of the particularly most dreadful attacks hence it causes an attacker to be able to divert the behavior of a network by just using a few copies of a previously hacked nodes \cite{conti2014clone}. Like the Sybil attack, the node-replication (clone) attack also can enable attackers to subvert data aggregation, misbehavior detection, and voting protocols by injecting false data or suppressing legitimate data \cite{parno2005distributed}.

\paragraph{Routing Attacks}
\begin{itemize}
	\item \textbf{Misdirection:} In misdirection attack, an attacker forwards ongoing messages to the wrong paths intentionally. This can be achieved by fabricating false routing advertisements and causing routing tables of the neighboring nodes' to update these false information \cite{wood2002denial}. This attack is also categorized as DoS attack, hence targeted nodes are blacked out completely and do not receive any further packets after the advertisement of the false routing information.
	\item \textbf{Network Partitioning:} A fully connected network is portioned to sub-networks in which the nodes in different sub-networks cannot communicate each other although they are connected.
	\item \textbf{Routing Loop:} A routing loop is introduced in a route path. It is created by spoofing routing updates. Suppose an adversary can determine that node A and node B are within radio range of each other. An adversary can send a forged routing update to node B with a spoofed source address indicating it came from node A. Node B will then mark node A as its parent and rebroadcast the routing update. Node A will then hear the routing update from node B and mark B as its parent. Messages sent to either A or B will be forever forwarded in a loop between the two of them. This leads to energy depletion and eventual node/network failure \cite{karlof2003secure}.
	\item \textbf{Rushing:} When this attack is performed against ``on-demand ad hoc network routing protocols'', it results in DoS in the network. For instance; AODV, DSR, and more secure protocols based on these, i.e.  ARAN, SAODV, and Ariadne, are not capable of discovering routes that are longer than two-hops when they are subjected to this kind of attack. The rushing attack is especially harmful to networks hence it can be executed by relatively weak adversaries \cite{hu2003rushing}.
	\item \textbf{Spoofed, Altered or Replayed Routing Information:} Routing information exchanged among nodes can be altered by malicious nodes to have a detrimental effect on the routing scheme.
\end{itemize}

\paragraph{RPL Exploit}
IoT consist of devices that are limited in resource like battery powered, memory,  processing capability, etc. For this kind o networks, a new network layer routing protocol is designed called RPL (Routing Protocol for Low-power and lossy networks) \cite{winter2012rpl}. RPL is light weight and does not have the full functionality of traditional routing protocols. RPL was proposed especially for data-sinks (multi-point to point communications) and is being adopted by IoT recently. Many attacks against the RPL of IoT have been presented in \cite{pongle2015survey}. 

As discussed and proposed in our paper, IoT is also susceptible to most of the attacks against WSNs. The attacks presented in \cite{pongle2015survey} and \cite{nawir2016internet} support this idea in a sense that other than a few attacks (which are specifically against RPL protocol), all of the attacks are same as the ``Attacks against Routing Layer'' presented in this section. These specific attacks against RPL protocol are: Local-Repair attack, Rank attack , DODAG version attack, DIS attack, and finally Neighbor attack. \cite{pongle2015survey}

In local repair attack, an attacker intentionally and periodically sends the
local repair message which is originally used for improvement of the link quality. This causes the neighboring nodes to go into local repair cycle. This
attack creates more impact on delivery ratio than any other kind of attack, generates more control packets and increases the end to end delay \cite{pongle2015survey}. 

DODAG stands for Destination Oriented Directed Acyclic Graph, which is created by RPL by forming a loop-free topology. DODAG organizes nodes in a hierarchical manner as single root, children and their further descendants \cite{mayzaud2014study}. 

In RPL, rank value increases from root to child node. In RPL DODAG rank attack, an attacker can exploit DODAG version system by advancing its rank in the hierarchical tree and gaining many children which are forced to route packets through the attacker parent. So, by intentionally changing the ranking value of itself, an attacker can attract many child nodes for selecting it as parent, and thereby attracts large traffic going toward the root node (main branch) to flow through itself.

Another exploit of  RPL DODAG version system is called DODAG version attack. It is executed by publishing a higher version number of the DODAG tree. When nodes receive the new higher version number in the  DODAG Information Object (DIO) messages, they start forming a new DODAG tree. This can cause the generation of new un-optimized topology and will bring inconsistencies in the network topology. The loops and rank inconsistencies created by the attack are generally located around the neighborhood of the attacker. \cite{pongle2015survey}

In  DODAG Information Solicitation (DIS) attack, an attacker sends DIS messages with fake IP addresses which causes the recipient node to re-generate the DIO messages, which eventually increases the overhead.

In RPL neighbor attack, a malicious node broadcast DIO messages that it received without adding information of itself. The node which receives this type of message may think that new neighbor node is sending this DIO message. The victim node tries to change the routing tables so that the pointed node is also included. This attack is somehow similar to the selective forwarding attack in which DIO messages selected only. This attack affects network by slightly increases the end to end delay, change in network topology, and some control overhead. However, it might have serious consequences when combined with other attacks.

\paragraph{6LoWPAN Exploit}
6LoWPAN is an Internet protocol devised for the IoT for the sake of extended usage of IPv6 by the smart-things. 6LoWPAN integrates IP-based infrastructures and WSNs by specifying how IPv6 packets are to be routed in constrained networks such as IEEE 802.15.4 networks by  fragmentation and reassembly of datagram data fields.

A specific attack for 6LoWPAN is fragment duplication attack, in which an attacker puts his own fragments in the fragmentation chain. The fragment duplication attack takes the advantage of the fact that a recipient cannot verify at the 6LoWPAN layer if a fragment originates from the same source as previously received fragments of the same IPv6 packet. Hence no authentication mechanism exists on the receiver at the time of reception for checking whether received fragment is an original or spoofed duplicate fragment, this attack can easily fool the receiver. The receiver cannot distinguish legitimate fragments from spoofed duplicates. Instead, it has to process all fragments that appear to belong to the same IPv6 packet according to the sender's MAC address and the 6LoWPAN datagram tag. Thus, an attacker can pretend as a legitimate node and exploit this weakness to engage in further attacks such as a DoS attack. \cite{pongle2015survey, hummen20136lowpan}. 

\paragraph{Sybil Attack}
A single node presents multiple identities to other nodes of the network. This causes confusion in the network; nodes receive contradicting routing paths that are passing through the attacker. This reduces the effectiveness of fault-tolerance schemes and poses a significant threat to geographic routing protocols \cite{gupta2015geographic}. Apart from these services it may also affect the performance of other schemes such as misbehavior detection, voting-based algorithms, data aggregation, fusion and distributed storage \cite{douceur2002sybil}.

\subsubsection{Attacks towards Transport Layer}
The transport layer of the OSI protocol stack manages end-to-end connections of the two nodes. At the transport layer, attacks exploit the protocols that maintain connection information at either end.\cite{raymond2008denial} All transport layer attacks are categorized and described as follows:

\paragraph{De-Synchronization} An attacker disrupts actual links among two nodes by de-synchronizing the transmissions in between them. An example of this type of attack is sending fabricated messages, such as faulty flag kind of sequences, (by transmitting forged packets with bogus sequence numbers or control flags that desynchronize endpoints so that they will re-transmit the data \cite{raymond2008denial}) continuously to both sides of the communicating parties, as a result, to force them losing their synchronization \cite{wood2002denial,znaidi2008ontology}.

\paragraph{MQTT Exploit}
The Message Queue Telemetry Transport (MQTT) \cite{iso2019mqtt} is a lightweight publish-and-subscribe connectivity protocol aimed at working on resource-constrained devices such as low power embedded sensors to enable them communicating. In the IoT context, MQTT is widely used to enable the communication between devices using a publish-and-subscribe messaging approach. However, MQTT does not include security layer by default and it is the user's responsibility to address security issues. In this direction, it is suggested to enable security for MQTT by issuing SSL/TLS with certificates and session key management. That is being said, owing to the multitude of heterogeneous devices, storing and managing the certificates and key exchanges for every session of IoT is burdensome. Furthermore, SSL/TLS can suffer from attacks such as BEAST, CRIME, RC4, and Heartbleed. Thus a scalable, lightweight and robust security mechanism is required for MQTT and its variants for deployment in IoT.\cite{singh2015secure}

\paragraph{Session Hijacking}
In computer science, this attack is referred as the ``exploitation of'' and ``tampering with'' a valid communication session (which is also called as session key) to gain unauthorized access on information or services of a system. As being an extension of IP networks, session hijacking of TCP messages will also be affecting and troublesome for IoT networks.

\paragraph{SYN-flooding} In a flooding attack, an attacker aims at exhausting the energy and/or the memory of a node, by flooding it with spurious messages. This is achieved, for instance, by sending multiple connection requests without ever completing the connection, thus overwhelming the buffer and eventually causing the node to be dead \cite{wood2002denial,znaidi2008ontology}.

More specifically, in a TCP SYN (synchronize) flooding attack, an adversary sends multiple TCP connection requests without ever completing the connection, thus overwhelming the target's half-open connection buffer. \cite{raymond2008denial}

\subsubsection{Attacks towards Application Layer}
Application layer protocols can also be exploited by DoS attacks. Protocols like node localization, time synchronization, data aggregation, association, and fusion can be cheated or hindered. For example, a malicious node that impersonates a beacon node and gives false location information or cheats with regard to its transmission power, i.e. transmitting with less or more power than it is supposed to do, may hamper the node localization scheme. Since this kind of attack diminishes the related network service, they can also be categorized as DoS attacks. An example of application layer DoS attack is path-based DoS attack and will be described below. All application layer attacks are categorized and described as follows: 

\paragraph{CoAP Exploit}
Constrained Application Protocol (CoAP) \cite{shelby2011constrained} is an application layer protocol designed as a replication of the HTTP for the small devices of IoT to provide communication ability with the rest of the Internet. Recently, many implementations of IoT are using CoAP, which indicates that it will have a crucial role in the future of IoT applications. 

As mentioned in \cite{rahman2016security}, there are several challenges related to security by the introduction of CoAP. It does not translate full functionality of HTTP, which creates security problems for multicast messages.

\paragraph{False Data Injection}
In order to influence the overall result of a measurement or a reading, captured nodes intentionally inject false data in the WSN. Therefore, it can be stated that this attack happens in a semantic level, hence it does not affect anything but the logic.

\paragraph{Path-based DoS}
As the name implies, this attack is a DoS attack that happens in the application layer. In this attack; an attacker overwhelms nodes but this time from long distance by again flooding an end-to-end communications path with either fabricated packets or replayed packets \cite{deng2005defending}. So, as a result, all the nodes along the path, from the source to destination (attacker to the base station) are affected by this attack.

\paragraph{Re-programming}
Once in a while, every network element either needs to be patched or re-programmed for version control, code acquisition, encoding-decoding, or when switching to a newly written program. That's also true for WSNs and IoT. If this reprogramming (or patch management per say) schedule is not kept secret, then adversaries can take advantage of this vulnerable time of the network by simply sending bogus messages to the nodes and pushing them into unstable or dead state \cite{ghildiyal2014analysis}.

\paragraph{Sensor Overwhelming}
Attacking or altering the sensitivity of the sensor measurements. Targeting sensors with spurious interference or completely overwhelming them with fake messages and inundating them with false stimuli.

\section{Defending Against Various Attacks towards the WSNs and IoT}\label{sec:defence}
Routing protocols can be designed such that an adversary cannot compromise nodes/messages or make the routing scheme dysfunction. This is the most effective approach with respect to the cost of the security scheme and effectiveness in defense of WSNs against the threats. Therefore, most of the techniques fall into this category. 

Preventive approaches are designed to counter known threats and may not be effective against new threats. Detection schemes for misbehaving or malfunctioning nodes can be designed in a more generic fashion. On the other hand, they can be more costly than preventive approaches. Finally, routing can be designed such that it still delivers the data packets to the destination when there is an attack. Such resilient techniques are also costly. 

Following subsections provide solutions (strategies and techniques) to defend (detect, prevent or mitigate) against various attacks towards the WSNs on all of the layers of the OSI protocol stack:

\subsection{Defense against Passive Attacks}
\subsubsection{Defense against Passive Information Gathering}
The communications in WSNs are achieved through the air, and therefore we do not know whether the packets arrive the intended people only or not. Hence detection of eavesdropping is almost impossible.

Link layer encryption would prevent outsider attacks such as eavesdropping, and some of the solutions are provided in \cite{karlof2003secure,perrig2002spins, deng2003performance, di2003lkhw,tubaishat2004secure}. The bulk of the external attacks towards WSNs can be avoided by link-layer encryption and authentication by using globally-shared keys. For example, to provide link layer encryption, Karlof \textit{et al.} \cite{karlof2004tinysec} proposed TinySec for WSNs. SNEP (Secure Network Encryption Protocol) under the SPINS \cite{perrig2002spins} protocol set, is also one another famous encryption protocol that is devised for WSNs.

In \cite{slijepcevic2002communication}, authors proposed SensorWare communication multicast model in which 3 different levels of link-layer encryption are provided by using the RC6 algorithm. They have chosen the RC6 algorithm, because its selection of the number of rounds parameter, has a direct effect on the security level of the algorithm. An initial set of master keys are shared by all sensors in the network. At any time, one of the master keys in the list of the master keys is active. The reason is, in order to expose less data for known-ciphertext attacks, more keys will be necessary as the lifetime of the network extends.  This selection of the keys is executed pseudo-random way, every node uses the same seed for randomizing function. Then, the random number is coupled to the list of the master keys to obtain the active master key. For the rest of the security model, the required keys for the 3 different levels of security are gathered from this active master key.

Random key pre-distribution schemes \cite{du2005pairwise, chan2003random, eschenauer2002key} help link layer encryption schemes by distributing the keys needed by the encryption algorithms, hence they help WSNs to protect information in transit and prevent eavesdropping, data and information spoofing.

\subsection{Defense against Active Attacks}
In Table~\ref{tab1}, some of the solutions \cite{agah2007preventing} to defend WSNs against various DoS attacks are summarized. However, this presentation is too generic and that's why we needed to present Table~\ref{tab2} in an effort to give the whole picture with a detailed categorization.

\begin{table}
	\caption{Solutions to defend WSNs against DoS attacks}
	\centering
	\setlength{\tabcolsep}{3pt}
	\begin{tabular}{|p{70pt}|p{155pt}|}
		\hline
		\textbf{DoS attack}& 
		\textbf{Defense strategy}\\[0.2ex]
		\hline
		\hline
		Radio interference & Usage of spread-spectrum communication\\[0.2ex]
		Physical tampering & Usage of tamper-resistant nodes\\ [0.2ex]
		Denying channel & Usage of error correction codes\\[0.2ex] 
		Blackhole & Usage of multiple routing paths\\[0.2ex] 
		Misdirection & Usage of source authorization\\[0.2ex] 
		Flooding & Limiting the total number of connections\\ 
		\hline
	\end{tabular}
	\label{tab1}
\end{table}

\subsubsection{Defense on Physical Layer}
In \cite{muraleedharan2006cross}, authors propose a cross-layer security mechanism, namely ``Swarm Intelligence'', to detect Jamming-DoS attacks. They also provide countermeasures to mitigate this kind of attack.
JAM $-$ a jammed area mapping service \cite{wood2003jam}, is proposed for detecting jamming DoS attack against WSNs. JAM provided a mapping protocol to detect a jammed region in a WSN. Besides, in terms of mitigation, JAM avoided the jammed part of the WSN by re-routing the packets thus provided a remedy for the Jamming DoS attack. 

In \cite{cagalj2007wormhole}, authors proposed wormhole technique, normally known to be an attack against WSNs, which could be effectively used to defer Jamming DoS attack against WSNs.  

In order to defend WSN from node tampering attacks, nodes might be equipped with tamper-resistant hardware, in which any kind of tamper attempt would wipe out the memory (and also any other data storage) so that confidential information (such as the secret keys) would not be leaked. Although this is a wise and excellent solution, it comes with an additional cost of hardware which would increase the total cost of the WSN installation and cripple the most appealing feature of such networks: ``low cost''. Other ways of fighting against node tampering would be disabling JTAG interface of the sensors and the use of good password protection for the bootstrap loader of the sensor boards \cite{znaidi2008ontology}. Finally, a na\"{\i}ve way of protecting the nodes against this kind of attacks would be simply hiding the nodes by camouflaging \cite{sharma2010wireless}. 

Detection of the tampering attempts would require routinely physical checking of the sensor nodes by eye or with special equipment such as magnifiers. Though, this task might be tedious considering the fact that sometimes nodes are deployed across dangerous and hard to reach places such as flooded zones, nuclear leakage areas, etc. 

\subsubsection{Defense on Data Link Layer}

\paragraph{Defense against Collusion and Exhaustion}
In order to defend WSNs against collusion and exhaustion attacks, request rate of each node might be limited to some certain value (by decreasing the MAC admission control rate), so that the network can discard extra requests from the same node (attacker) \cite{znaidi2008ontology}. Another solution would be the employment of time division multiplexing (TDM) technique which would provide dedicated time slots to each node to transmit their packets. This would allow each node to have a tiny period of time to access the channel. By this way, the channel usage of each node is limited and attacks related to channel abusing are prevented. If the corruption of the packets occurs partially (in bursts), employment of error detection and error correction codes would be beneficial tools to fight against this kind of attacks \cite{sharma2010wireless}.

\paragraph{Defense against Denial of Sleep and Link Layer Flooding}
Liu et al.'s anomaly detection-based IDS proposal works on ad hoc networks and therefore fits well to the WSNs and IoT. In their proposal, each node participates to the detection of the abnormal nodes. Packet traffic at MAC layer is analyzed and evaluated. Distributed and cooperative anomaly detection is achieved by creating feature vectors at each node and then fulfilling cross-feature analyses. This is resulted with local or global response (either response from single node or a collaborative response from multiple nodes) according to the seriousness of the attack situation. The proposed IDS has shown to be effective against MAC layer attacks, especially on denial of sleep and flooding attacks. 
\cite{liu2005mac}

\paragraph{Defense against De-synchronization}
Many applications require the IoT to have a low and deterministic delay, especially for IIoT applications where delay tolerance is low.  Having a deterministic network delay in a WSN is challenging in the best case when traditional MAC protocols are employed. To enable low power, high reliability and deterministic WSNs, IETF recently proposed 6TiSCH protocol that uses time slotted channel hopping (TSCH) MAC with IPv6 addressing. This protocol dynamically assigns bandwidth resources to the nodes in the network according to the application requirements, hence it provides a communications stack for low power and lossy networks such as IoT. The proposal includes secure communication of MAC layer frames so that any kind of eavesdropping or packet capturing would have no benefit of these.\cite{accettura2014optimal}

\paragraph{Defense against Unfairness} 
In \cite{wood2002denial}, usage of small frames is offered as a defense solution to this kind of attack. By this way, an attacker would capture the channel only for a small period of time and unfairness would be avoided.
\subsubsection{Defense on Network Layer}
\paragraph{Defense against Blackhole}
Liu et al.'s anomaly detection-based IDS proposal works on each node and have also shown to be effective (but less compared to detecting attacks against MAC layer) on network layer, especially for detection of blackhole and grayhole (packet-dropping) attacks. \cite{liu2005mac}

In \cite{karakehayov2005using}, authors proposed REWARD scheme, in order to detect blackhole attacks against WSNs. Besides, this scheme provides a routing algorithm, in which detected blackholes are avoided hence the WSN resumes normal operation under blackhole attack. 

In \cite{krishnan2016qos}, authors discussed the effects of network topology selection in WSNs on blackhole attacks. Their findings suggest that mesh topology is more resilient against blackhole attacks compared to star and tree topologies. 

In \cite{prathapani2009intelligent}, Prathapani \textit{et al.} proposed the use of Honeypots, which are strolling software agents that create dummy Route-Request (RREQ) packets to attract and catch blackhole attack performers. Authors have illustrated the positive performance of the proposed detection approach by extensive simulation results using the ns-2 simulator.

Watchdog based intrusion detection system proposed in \cite{tiwari2009designing}, is efficient in detecting blackhole and selective forwarding attacks. Sensor nodes in the clustered-WSN watch their neighbors and collaborate with the head of the cluster in order to be able to detect misbehavior. Although sensor nodes do not have an exhaustive view of the network, still they can be effective in detecting intrusions with some definite probability and report it to the head of the cluster.

In \cite{medadian2009combat}, Medadian \textit{et al.} presented a method to cope with the blackhole attacks which introduces negotiating with neighboring nodes that claim to possess a route to the destination. If a node receives an RREP packet (corresponding answer to an RREQ packet) then it forwards this packet to the source and starts an evaluation procedure about the replier. The evaluation procedure is based upon opinions of the nodes about the replier. All activities of a node are saved by its neighbors. Whenever the neighbors of a particular node are asked to send their opinions about it, requester node gathers all opinions of the neighbors and concludes about the maliciousness of the replier node. The conclusion is based on several rules. According to authors' simulation results, it has shown that the presented MAODV protocol provided not only better security but also performed better in terms of packet delivery ratio than the legacy AODV protocol in the existence of blackholes with minimum added delay and overhead.

In ActiveTrust \cite{liu2016activetrust}, in order to detect blackhole attacks in WSNs, authors have created multi-detection routes in the areas which have remaining energy. Hence an adversary does not know the detection routes, (s)he will target those and while executing the action (s)he will be disclosed. By this method, location and behavior of an adversary can be determined by ActiveTrust to be used for avoiding blackholes during the creation of the final data routes.

In \cite{wazid2013detection}, authors presented a technique for detecting and preventing blackhole attacks in clustered-WSNs. In this methodology, a node called coordinator is selected by all nodes dependent on the cluster-head election criterion. Coordinator node has the responsibility in the authentication process, detection of intermediate node failures as well as blackhole attacks. If a blackhole attacker is detected in the cluster, then the coordinator removes it from the cluster, hence all the communications with this specific node are terminated.

Misra \textit{et al.} \cite{misra2011bambi} proposed a method called BAMBi, which is based on the deployment of spatially diverse base stations in the WSN to cope with the consequences of having blackholes on the data transmission. Presented method demands transmission of extra copies of a packet from each node to all base stations. To ensure that every node has a route to it, each base station uses TinyOS beaconing. The beacon packet from any BS consists of: the ID of the sender of the packet, the ID of the base station from which it originated, and the hop count of the sender from the base station to the node. This beaconing process creates a routing tree in the network and used for detection and mitigation of blackhole attacks against the network. According to their simulation results, authors have shown that the proposed method has achieved over $99\%$ success rate in packet deliveries in the presence of blackhole attack. 

In \cite{amouri2015simple}, Amouri \textit{et al.} proposed a framework of intrusion detection for MANETs. The framework considered a hierarchical architecture where the intrusion detection is distributed through a set of promiscuous zones (PZ). Unlike the traditional approach where the nodes are promiscuous all the time, in their scheme the nodes are promiscuous for the period that they are in the PZ. Once a node leaves the virtual PZ its promiscuity is turned off in order to save energy. Authors have used a C4.5 decision tree to learn the network behavior under blackhole attack, and after exhaustive evaluation, authors have shown that their approach was able to recognize blackhole attacks with up to $97\%$ accuracy.

\paragraph{Defense against HELLO Flooding}
One possible solution to this problem is provided in \cite{karlof2003secure}: Force every node to authenticate each of its neighbors with an identity verification protocol using a trusted base station. If the protocol sends messages in both directions over the link between the nodes, HELLO floods are prevented when the adversary only has a powerful transmitter because the protocol verifies the bi-directionality of the link.

In \cite{hamid2006routing}, authors presented a shared-secret method based on probability to defend against HELLO flooding attack. The proposed method has 2 steps in terms of defense: 1) Bidirectional verification technique to detect HELLO floods. 2) ``Multi-path multi-base station routing'' in order to defer HELLO floods.

$\mu$-TESLA ``(Micro Timed Efficient Streaming Loss-tolerant Authentication) protocol'', under the SPINS \cite{perrig2002spins} protocol set, has been proposed to provide authenticated broadcasting in an effort to prevent HELLO flooding attack; and achieved this by employing symmetric key cryptography which required minimum packet overhead. 

\paragraph{Defense against Node-Replication (Clone)}
In \cite{zhang2006location} and \cite{duan2011efficient}, authors proposed location based Public Key Cryptography (PKC) algorithms in order to prevent clone attacks. Every sensor node has a private key associated with the location of itself. These location-based keys are employed to prevent node replications. 

In detecting clone attacks, there are a variety of solutions provided. More interested readers would refer to \cite{zhu2012detecting}, hence detailed comparisons of the schemes proposed to detect node-replication attacks are provided in. Here, they are summarized under 2 categories:

\textbf{Centralized Solutions:} 
Conventional methods to detect a node replication attack usually include centralized computing based on node locations or the number of simultaneous connections, which is vulnerable to the single-point failure \cite{parno2005distributed}. Mostly, centralized detection schemes do not have significant advantages over distributed detection schemes; and bear similar deficiencies (such as any compromisation at the base-station would transform the provided-solution to an ineffective one).

\hspace{0.1cm} Choi \textit{et al.} \cite{choi2007set} presented a scheme to detect centralized node-replication attacks and named it as ``SET''. SET aims at reducing the overhead caused by the detection, with computation set of operations (``intersection'', ``union'') related to exclusive subsets of the WSN. Not only SET protocol is very complicated to implement (owing to complex components like interleaved authentication and authenticated subset covering), but also has an increased overload.

\hspace{0.1cm} Brooks \textit{et al.} \cite{brooks2007detection} presented a protocol to detect clone attacks based on ``random pairwise key pre-distribution schemes''. Proposed scheme addressed the problem of detecting replicated cryptographic keys instead of the replicated sensor nodes. By analyzing statistical data from the sensor node authentications, the proposed scheme detects the replicated keys as follows: The keys that are exceeding a predefined usage count (threshold) are declared as replicated and revoked from the WSN. 

\hspace{0.1cm} Xing \textit{et al.} \cite{xing2008real} presented a method to detect node-replication attacks with the encoded information related to nodes' community network called the ``social fingerprint''. In this method, nodes collaborate to create each other's fingerprints then send these data to the base station for further conclusion process. However, in the proposed scheme, the expected number of nodes is not adjustable; therefore not only node addition but also node revocation cannot be managed, causing the flexibility of the WSN to disappear. 

\hspace{0.1cm} Ho \textit{et al.} \cite{ho2009fast} presented an efficient and fast node-replication detection model for mobile WSNs called 	``speed test''. In this model, the base-stations compute the instant maximum speed of a node, and denote it as $Vmax$ and then compare it with the system configured maximum theoretical speed $Vmax-T$. An authentic node should be never moving faster than $Vmax-T$. If the instant-measured $Vmax$ is to be found above the configured $Vmax-T$, then it is most probable that there are at least two sensor nodes using the same ID in the WSN, the indication of a replication attack. Although the proposed scheme sounds very appealing, it might introduce errors causing high false-negative and false-positive outcomes due to the synchronization errors.

\hspace{0.1cm} Butun \textit{et al.} \cite{butun2015intrusion} proposed an IDS system for clustered-WSNs based on multi-level clustering. In their approach, all the member nodes of each cluster share secret keys with their cluster head. Each node is specific to their unique cluster and cannot be used in any other cluster elsewhere in the network. Therefore, any attempt of node replication attack is easily detected and prevented by using this kind of clustered approach.

\textbf{Distributed Solutions:}
In \cite{parno2005distributed}, Parno \textit{et al.} presented a method for distributed detection of the clone attacks. On this approach, each sensor node is considered to know its own location, and it is mandatory to send this location information to a set of watchdog sensors. If a watchdog sensor detects an abuse in the location declarations of a sensor node, this node then becomes as suspicious of having a replicated (once or more) identity in the network. In order to cope with this problem and to certify the authentic location claims of the nodes, public-key cryptography is employed.

\hspace{0.1cm} Another distributed detection method of clone attacks is devised by Conti \textit{et al.} \cite{conti2014clone}. In their proposal, ``HIP-HOP (History Information-exchange Protocol and its optimized version)'' protocol is devised. Proposed protocols make use of local neighborhood (one-hop) communications and mobility of nodes, and can be considered as light-weight for the amount of computation required. While detecting clones, the proposed protocols work in an effective, distributed and cooperative way.

\hspace{0.1cm} In node-to-network broadcasting (N2NB) scheme \cite{parno2005distributed}, the whole network is flooded by every node with authenticated-broadcast messages. This is in an effort of each node by claiming its own location. In the pre-condition that the broadcast messages reach each node, the N2NB protocol is reported to achieve $100\%$ detection rate.

\hspace{0.1cm} In deterministic multicast (DM) scheme \cite{parno2005distributed}, the aim of the design is to reduce the communications cost, and the main goal is sending location claim of a node only towards a certain number of nodes, chosen in a deterministic way, to serve as watchdogs in the network.

\hspace{0.1cm} ``Randomized multicast (RM)'' and ``line-selected multicast (LSM)'' protocols are two probabilistic algorithms, both proposed by Parno \textit{et al.} \cite{parno2005distributed}. RM protocol dissipates node-location claims to a selected random number of watchdogs, leveraging ``combinatorics theory'' in detecting replicas; while LSM leverages the network topology related to routing to assign extra watchdogs for the claimer and employs ``geometric probability'' methodology for the detection.

\hspace{0.1cm} Zhu \textit{et al.} \cite{zhu2007efficient} presented 2 schemes: ``Single deterministic cell (SDC)'' and ``parallel multiple probabilistic cells (P-MPC)''. Both schemes are derived from DM, and can be employed as ``network-wide deterministic multicast'' protocol, followed by probabilistic storage and in-cell broadcast methodologies.

\hspace{0.1cm} Conti \textit{et al.} \cite{conti2007randomized} presented ``randomized, efficient, and distributed (RED) protocol''. This method combined both advantages of RM and DM. In the state of the art, RED is one of the most appealing clone-detection algorithms.

\hspace{0.1cm}In \cite{zhang2009memory}, Zhang \textit{et al.} proposed 4 clone-detection protocols with the name of ``memory efficient multicast (MEM)''.

\hspace{0.1cm} Li and Gong \cite{li2009randomly} proposed a simple form of N2NB named as ``randomly directed exploration (RDE)'', in which location claims along with the claimer's neighborhood list are sent so that each of the forwarding nodes on the path fairly constitutes a line. RDE is only feasible for ideal network model, hence detection rate might not be at an important level even for a convex deployment field.

\paragraph{Defense against Selective Forwarding (Grayhole)}
There are two approaches to defend against selective forwarding attacks:
\begin{itemize}
	\item Detecting the nodes that selectively forwarding.
	\item Developing routing schemes that are more resilient and can deliver packets even when there is a selective forwarding attack.
\end{itemize}

One approach in detecting the nodes that are selectively forwarding is based on acknowledgments \cite{yu2006detecting}. Every intermediate node that forwards a packet waits for an acknowledgment from the next hop. If the next hop node does not return the same number of acknowledgments as the number of packets sent, the node generates an alarm about the next hop node. However, compromised nodes can also generate acknowledgments for the packets that they dropped, which make this scheme fail. 

Multi-path routing can be an effective way to mitigate selective forwarding and blackhole attacks \cite{karlof2003secure}. This requires at least link-disjoint paths, where two paths may share some nodes but no link. Of course, node-disjoint paths, where two paths do not have any node in common, are better and reduce the risk of selective forwarding attack compared to link-disjoint paths. However, disjoint paths are not always available, and when paths are not disjoint, if the selectively forwarding node is the node common to all the paths, then the attack can become as effective as in single-path routing. 

Braided paths \cite{ganesan2001highly} may have nodes in common, but have no links in common (i.e., no two consecutive nodes in common). The use of multiple braided paths may provide probabilistic protection against selective forwarding and use only localized information.

In \cite{hai2008detecting}, Hai and Huh proposed a lightweight detection algorithm to detect selective forwarding attacks in WSNs. The proposed algorithm is based on two-hop neighbor knowledge including two routing rules. The proposed algorithm has been evaluated for WSNs and has found to be effective even under high-density network conditions along with the high probability of collisions.

In \cite{brown2008detection}, authors describe an efficient scheme for reporting packet drops. They also present an effective scheme, namely ``Wald's Sequential Probability Ratio Test'', for detecting the selective forwarding attack in a heterogeneous sensor network. According to presented simulation results, proposed scheme not only achieves high detection ratio but also low false alarm rate.

Wang \textit{et al.} \cite{wang2003supporting} proposed a failure detection framework to detect the selective forwarding attack. The observation is that for a routing node, the number of packets it forwards must be equal to the number of packets it receives. In their framework, each sensor node can work under a promiscuous mode so that it can overhear the transmission of neighboring nodes. If a neighbor of a suspected node finds that the number of packets that the suspected node fails to forward exceeds a certain threshold, the neighbor can collaborate with other neighbors of the suspected node, and the opinions from the neighbors of the suspected node are collected to form a decision about the suspected node.

\paragraph{Defense against Sinkhole} 
An algorithm which detects sinkhole attacks is presented in \cite{ngai2006intruder}. Proposed algorithm first finds a list of suspected nodes, and then effectively identifies the intruder in the list through a network flow graph.

In Shafiei \textit{et al.}'s \cite{shafiei2014detection} proposal; two schemes in an effort of detecting sinkhole attacks are presented. The idea behind these schemes is that the sensor nodes in the vicinity of the sinkhole would deplete their energies more rapidly compared to other sensor nodes. This is due to the fact that, routes to the base station passing through sinkhole are supposedly more found to be more attracting thus are chosen more often. Therefore, an energy hole forms in the vicinity of each and every sinkhole. The first scheme has proposed that base stations to utilize a geo-statistical sampling methodology for detecting possible sinkholes in the entire area of the WSN using the energy spending pattern (remaining energies of each area) and issuing an estimator based on the data extracted from the statistics. Depending on the result of the estimator, base station commands all nodes to bypass the doubtful area (possible attack region) in their process of routing. The second scheme utilizes distributed monitoring methodology in order detecting areas having less average remaining energy level.

Zhang \textit{et al.} \cite{zhang2014sinkhole} presented a scheme for detecting sinkhole attacks towards WSNs based upon redundancy mechanism. In the proposed scheme; in order to detect sinkholes, messages are sent to the suspicious nodes through multi-paths. By evaluating the replied messages by the suspicious nodes comprehensively, the attacker nodes are confirmed.

Teng and Zhang \cite{teng2010sera} proposed ``SeRA: A secure routing algorithm against sinkhole attacks for mobile WSNs''. The proposed algorithm is based on the Tiny-AODV protocol. First, a couple of mobile-agents communicate each sensor node to gather the network connection data to construct the global data matrix of the sensor nodes. Then, through the SeRA routing algorithm, the sinkhole node can be efficiently avoided. 

\paragraph{Defense against Sybil}
In order detecting the Sybil-attack, two protocols are presented in \cite{newsome2004sybil}. The first protocol employs ``radio resource testing'' in which every sensor node declares a unique-channel to each of its neighbors. Then, it tests to see whether the neighbors communicate with itself through the pre-declared channels. Hence radio circuitry of a sensor node generally cannot handle simultaneous send and receive actions on more than one channel, the failure at the communications through one channel might be an indication of the Sybil-attack. The other protocol employs the ``ID-based symmetric keys''. For instance, there is a global pool of keys that each sensor node has a key associated to its ID, and every node is pre-loaded with these keys. The ID of a suspicious sensor node is examined by witness nodes based on keys shared in between the suspicious sensor node and the witness sensor nodes. 

In \cite{sarigiannidis2015detecting}, authors proposed ``rule-based anomaly detection system (RADS)'' in order monitor and detect Sybil-attack towards sensor networks on time. The presented method relied on ``ultra-wideband ranging-based detection algorithm''. RADS works in a distributed fashion in which every node have the ability to trigger alarms, in case they are suspicious of a node in their vicinity.  The presented method not only detects Sybil-attack in the absence of central management authority or a third party trusted-network-entity but also provides defending methods against the adversaries by using isolation technique towards both the attacker sensor node and the compromised Sybil-nodes.   

To defend against Sybil attacks, the identities of every node should be verified. This can be done either directly or indirectly. In the direct validation, a node directly verifies whether the identity of a neighboring node is valid. For example, a node may assign each of its neighbors a separate channel to communicate and ask them to transmit during a period. Then it checks these channels in a random order within that period. If a node is transmitting in its assigned channel, the node is a physical node. If no transmission is detected on a channel, it indicates that the node assigned to that channel may not be a physical node \cite{newsome2004sybil}.

In indirect validation method, another trusted node provides the verification for the identity of a node. For example, every node may share a unique key with the base station. When two nodes need to establish a link between them, they verify each other's identity through the base station by using these keys \cite{karlof2003secure}. At the same time, they can be assigned a session key. Nodes can also be allowed to establish links with a limited number of neighboring nodes. Thus, compromised nodes can only communicate with a limited number of verified neighboring nodes, which also limits the impact of Sybil attacks.

Moreover, ID-based public keys \cite{zhang2006location} also can defeat the Sybil attack because both the ID and location information were taken into the generation of key material during the initialization phase, hence multiple identities need multiple keys, and this is impossible for a malicious node to achieve.

\paragraph{Defense against Wormhole} 
Wormholes are difficult to detect because an adversary passes the packets to a distant point from the point at which they are received by using a single hop out-of-band channel. This channel cannot be listened to by the network. Moreover, the real copy of the packet reaches the point that receives the replayed copy later than the replayed copy. Therefore, the replayed copy is fresher than the real copy.

Detection mechanisms against wormhole attacks can be based on temporal and spatial analysis of the packets. To detect the Wormhole attack, Hu \textit{et al.} proposed to use packet leashes \cite{hu2003packet}, where location or timing information is embedded in packets, to limit the maximum range over which packets can be tunneled. They require that each node either knows its location or has a tightly synchronized clock so that this information can be used to calculate the maximum distance that a relayed packet could travel.

Directional antennas \cite{hu2004using} were also used to defend against the Wormhole attack, where some direction information is used to detect the replayed packets. However, these defenses target ad hoc networks and require expensive hardware devices, which may be infeasible for most resource-constrained sensor networks.

Wang and Bhargava \cite{wang2004visualization} proposed to use centralized computing to detect the Wormhole attack in sensor networks, in which a controller collects the location information for all nodes to reconstruct the network topology such that any topological distortion can be visualized. However, the visualization approach incurs too much communication overhead, especially when malicious nodes move around in the entire network because each location change of the Wormhole triggers a new round of execution of the topology reconstruction algorithm. 

Location-based keys \cite{zhang2006location} can effectively prevent the Wormhole attack because each packet is authenticated by the location-based key, which includes authentication information in accordance with the position information of the nodes. This method excludes the effects of wormholes by attesting the position of the nodes along with their location-based keys. Attacker nodes would simply be denied to join the network if their location-based keys are invalid.

Kaissi \textit{et al.} \cite{el2005dawwsen} proposed ``DAWWSEN'', a proactive routing protocol based on the construction of a hierarchical tree where the base station is the root node, and the sensor nodes are the internal or the leaf nodes of the tree. DAWWSEN fights against wormhole attacks by creating a hierarchical 3-way handshake routing tree and any attempt of creating wormholes are discarded by this generated routing tree.

\paragraph{Defense against 6LoWPAN Exploit}
In \cite{glissa20186lowpsec}, authors proposed a new security protocol referred to as 6LowPSec, providing a favorable end-to-end security solution that will be working on 6LoWPAN protocol. 6LowPSec employs existing hardware security features specified by the MAC security sublayer. However, the proposed solution functions at the adaptation layer and has not been evaluated by the research community yet.

A specific solution to 6LoWPAN fragment duplication attack is proposed by Hummen \textit{et al.}; the content chaining scheme uses cryptography to verify that received fragments belong to the same packet, so that spoofed (duplicated) fragments would be avoided \cite{hummen20136lowpan}.

\paragraph{Defense against RPL Exploit}
As RPL is a newly proposed networking technology, attacks towards it are being discovered these days. Hence, there are not many solutions proposed in defending it against those attacks. 

TRAIL (TRust Anchor Interconnection Loop) \cite{perrey2013trail} , is proposed as a generic scheme for topology authentication in RPL. It detects and prevents topological inconsistencies by enabling each node to validate its upward path to the root and to detect rank spoofing on it. TRAIL also minimizes network message exchanges and node resource consumption. Hence it protects RPL against \textit{rank attack}. 

As suggested in \cite{mayzaud2014study}, those newly developed defense solutions need to be as simple as possible considering that RPL works on low-power lossy networks.  Same authors suggest integrity check for the version field of the DIO messages to thwart \textit{DODAG version attack} against RPL.

To defend against RPL rank attacks, VeRA (Version number and Rank
Authentication) is proposed by Dvir \textit{et al.} \cite{dvir2011vera}. 
It is a security scheme that prevents misbehaving nodes manipulating their rank values for attack purposes. VeRA prevents publishing an illegitimate rank value by generating the hash chaining using random numbers chosen by the root node. An attacker cannot change the rank value as it requires the previous hash chain value to generate the new one. In order for a node to change its rank,  rank authentication is needed. That is provided by the root node consisting of MAC (Message Authentication Code) generated from the max rank hash value and next version number as key. VeRA security scheme also prevents DODAG version attack by providing verification to version number using digital signature and MAC.

To defend against RPL local repair attacks, inclusion of timer (with a long waiting time setting) to the local link repair requests might be a good solution. Of course, VeRA will also partially defend the network against this kind of misuse, as misbehaving nodes are excluded from the network.

\subsubsection{Defense on Transport Layer}
\paragraph{Defense against De-Synchronization}
As pointed out in \cite{wood2002denial}, one way of defending against this kind of attack is authenticating all of the packets being exchanged, inclusively entire control field existed in the transport protocol header. On the assumption that an adversary cannot forge authentication, the endpoints of the communication would detect and ignore any malicious packets.

\paragraph{Defense against SYN-Flooding}
Connectionless transport protocols are immune to SYN-Flooding attacks, but they might not provide the necessary transport-layer functionality to applications. One way of defense against this type of attack is SYN-cookies, which encode information from the client's TCP-SYN message and return it to the client to avoid maintaining state at the server. However, this technique's computational and message overhead makes it undesirable for WSNs and IoT \cite{raymond2008denial}. 

As stressed out in \cite{wood2002denial}, usage of client puzzles is a solution to defend against flooding attacks. The method presented in \cite{aura2000resistant} requires clients to demonstrate the commitment to solve ``client puzzles'' in each connection. The server or base station is dedicated to create and verify puzzles for each client. The Base station has the responsibility to dissipate the puzzles to the network members (nodes) and whoever wishing to connect to the base station needs to solve the puzzle beforehand. 
\paragraph{Defense against MQTT Exploit}
Neisse \textit{et al.} \cite{neisse2014enforcement} have proposed enforcement of security policy rules at the MQTT layer, which can be used to support security and privacy requirements. Performance results of their presented implementation show that enforcing complex security policies introduce an additional delay of 10 ms. However, in mission-critical applications such as traffic control or IIoT \cite{da2014internet}, any value above 1 ms may not be tolerable and can cause system failures or accidents. Therefore, in order to be applicable for those scenarios, the delay related performance of the proposed scheme needs to be improved.

Singh \textit{et al.} have proposed a Secure MQTT (SMQTT) protocol which augments security feature for the existing MQTT protocol and its variants based on lightweight Attribute Based Encryption (ABE) over elliptic curves. They take advantage of using ABE which supports broadcast encryption: With one cycle of encryption a message is delivered to multiple intended recipients and thus suitable for IoT applications \cite{singh2015secure}.

\paragraph{Defense against Session Hijacking}
Light-weight user authentication algorithm for optimized routing in mobile networks for defending against session hijacking attacks by Song \textit{et al.} \cite{song2009secure}.

\subsubsection{Defense on Application Layer}
\paragraph{Defense against CoAP Exploit}
As mentioned in \cite{rahman2016security}, the Transport Layer Security (TLS) variant for CoAP is Datagram Transport Layer Security (DTLS) - it is called CoAPs in this phase, which provides an additional  protection layer. However, this comes with a heavy cost of additional computation and big amount of handshake content in the message, which causes message fragmentation.

\paragraph{Defense against False Injection}
Ye \textit{et al.} \cite{ye2005statistical} presented ``statistical en-route filtering (SEF)'' technique for detecting false information (data) injected in the WSN. Besides, as a mitigation, a methodology is provided in order to filter those injected data and how to achieve a consensus on a measurement by introducing a collective secret.

\begin{table*}
	\caption{Layered categorization of cyber-security attacks towards WSNs and IoT along with the proposed solutions to defend against those attacks.}
	\centering
	\setlength{\tabcolsep}{3pt}
	\begin{tabular}{|p{71pt}|p{34pt}|p{188pt}|p{188pt}|}
		\hline
		%& & &\\
		\textbf{Attack type}&\textbf{Layer}&
		\textbf{Proposed Solutions for Detection} &
		\textbf{Proposed Solutions for Prevention/Mitigation}\\
		%& & &\\
		\hline
		\hline		
		%& & &\\
		Eavesdropping &	All layers	& N/A &	Link-layer encryption \cite{karlof2003secure,perrig2002spins, deng2003performance, di2003lkhw,tubaishat2004secure,karlof2004tinysec}, SensorWare communication multicast model \cite{slijepcevic2002communication}, 
		Key pre-distribution \cite{du2005pairwise, chan2003random, eschenauer2002key}\\
		\hline
		\hline
		%& & &\\
		Jamming-DoS & Physical & Swarm intelligence \cite{muraleedharan2006cross}, 
		JAM (mapping) \cite{wood2003jam} & 
		Usage of spread-spectrum communication \cite{agah2007preventing}, JAM (re-routing) \cite{wood2003jam}, Wormhole technique \cite{cagalj2007wormhole}\\
		& & &\\
		Tampering & Physical &	Routinely executing physical checks & Tamper resistant hardware, disabling JTAG and/or protecting bootstrap loader \cite{znaidi2008ontology}, camouflaging \cite{sharma2010wireless}\\
		%& & &\\
		\hline
		\hline
		%& & &\\
		Collusion and Exhaustion & MAC & Error detection codes \cite{sharma2010wireless} & TDM, Error correction codes \cite{sharma2010wireless}\\
		& & &\\
		Denial of sleep and flooding & MAC & Anomaly detection on motes \cite{liu2005mac} & N/A\\
		& & &\\	
		De-Synchronization&MAC&N/A&6TiSCH \cite{accettura2014optimal}\\
		& & &\\		
		Unfairness & MAC & N/A & Usage of small frames \cite{wood2002denial}\\
		%& & &\\
		\hline
		\hline
		%& & &\\
		Blackhole&Network& Anomaly detection on motes \cite{liu2005mac}, REWARD \cite{karakehayov2005using}, ActiveTrust \cite{liu2016activetrust}, Packet count \cite{wazid2013detection}, TinyOS beaconing \cite{misra2011bambi}, Honeypot \cite{prathapani2009intelligent}, Watchdog \cite{tiwari2009designing}, Pseudo clustering algorithm \cite{amouri2015simple} 
		& REWARD routing \cite{karakehayov2005using}, Multi-path routing \cite{karlof2003secure,agah2007preventing,ganesan2001highly}, Mesh network topology \cite{krishnan2016qos}, ActiveTrust routing \cite{liu2016activetrust}, Isolation \cite{wazid2013detection}, BAMBi \cite{misra2011bambi}, MAODV \cite{medadian2009combat}\\		
		& & &\\
		HELLO flooding&Network&Bidirectional verification technique \cite{hamid2006routing}&Identity verification protocol \cite{karlof2003secure}, Multi-path multi-base station routing \cite{hamid2006routing}, $\mu$-TESLA \cite{perrig2002spins}\\
		& & &\\
		Node-Replication (Clone)&Network&\textit{Centralized solutions:}
		SET \cite{choi2007set}, Random pair wise key pre-distribution \cite{brooks2007detection}, Social fingerprinting \cite{xing2008real}, Speed test  \cite{ho2009fast}, Multi-level clustering \cite{butun2015intrusion}& ID-based public keys \cite{zhang2006location}, Location-based key management \cite{duan2011efficient}, Multi-level clustering \cite{butun2015intrusion}\\		
		%& & &\\
		& &\textit{Distributed solutions:}
		HIP-HOP \cite{conti2014clone}, N2NB, DM, RM and LSM \cite{parno2005distributed}, SDC and P-MPC \cite{zhu2007efficient}, RED \cite{conti2007randomized}, MEM \cite{zhang2009memory}, RDE \cite{li2009randomly}&	\\
		& & &\\
		RPL DODAG version&Network&N/A&Integrity check \cite{mayzaud2014study}, VeRA  \cite{dvir2011vera}\\	
		& & &\\
		RPL local repair&Network&N/A&Inclusion of timer in local link repair messages, VeRA  \cite{dvir2011vera}\\
		& & &\\
		RPL rank&Network&N/A&TRAIL \cite{perrey2013trail}, VeRA  \cite{dvir2011vera}\\
		& & &\\
		Selective forwarding (Grayhole)&Network& Anomaly detection on motes \cite{liu2005mac}, Acknowledgment monitoring \cite{yu2006detecting},Neighbor knowledge \cite{hai2008detecting}, Reporting packet drops \cite{brown2008detection}, Failure detection framework \cite{wang2003supporting}, Watchdog \cite{tiwari2009designing} &	Multi-path routing \cite{karlof2003secure,ganesan2001highly}, Usage of source authorization \cite{agah2007preventing}\\
		& & &\\
		Sinkhole&Network&Network flow graph \cite{ngai2006intruder},Geo-statistical sampling approach and distributed monitoring approach \cite{shafiei2014detection}, Redundancy mechanism \cite{zhang2014sinkhole} &Secure routing algorithm \cite{teng2010sera}\\		
		& & &\\
		6LoWPAN exploit&Network&N/A&6LowPSec \cite{glissa20186lowpsec}, Content chaining scheme \cite{hummen20136lowpan}\\		
		& & &\\
		Sybil&Network&Radio resource testing, ID-based symmetric keys, registration, position verification, code attestation \cite{newsome2004sybil}, RADS \cite{sarigiannidis2015detecting} &Indirect validation \cite{karlof2003secure}, Identity verification \cite{newsome2004sybil},  Isolation \cite{sarigiannidis2015detecting},	ID-based public keys \cite{zhang2006location}\\		
		& & &\\
		Wormhole&Network&Packet Leashes \cite{hu2003packet}, Directional antennas \cite{hu2004using}&Location-based keys \cite{zhang2006location}, Centralized computing \cite{wang2004visualization}, DAWWSEN \cite{el2005dawwsen}\\		
		%& & &\\
		\hline
		\hline
		%& & &\\
		De-Synchronization&Transport&N/A&Usage of authentication including transport layer protocol headers \cite{wood2002denial}\\
		& & &\\
		SYN-flooding&Transport&N/A&SYN-cookies \cite{raymond2008denial}, Client puzzles \cite{aura2000resistant}\\
		& & &\\
		MQTT exploit&Transport&N/A&Enforcement of security policies \cite{neisse2014enforcement}, SMQTT \cite{singh2015secure}\\
		& & &\\
		Session hijacking&Transport&N/A&Light-weight user authentication algorithm for optimized routing in mobile networks \cite{song2009secure}\\
		\hline
		\hline
		CoAP exploit&App.&N/A&CoAPs, employment of DTLS \cite{rahman2016security}\\
		& & &\\
		False data injection&App.&SET \cite{ye2005statistical} &Collective secret \cite{ye2005statistical}\\
		& & &\\
		Path-based DoS&App.&N/A&One-way hash chains \cite{deng2005defending} \\	
		%& & &\\
		\hline
	\end{tabular}
	\label{tab2}
\end{table*}

\paragraph{Defense against Path-Based DoS}
To defend against Path-based DoS (PDoS) attacks, Deng \textit{et al.} presented a light-weight, efficient and robust method which employed ``one-way hash chains'' that allowed intermediate sensor nodes to detect replayed and spurious packets \cite{deng2005defending}. In general, PDoS attack targets at intermediary sensor nodes in a ``multi-hop end-to-end data path'' of sensor networks. In the proposed solution of Deng \textit{et al.} \cite{deng2005defending}, a ``one-way hash chain'' in each sensor node towards a communications path is configured which enabled each intermediary sensor node ability in the detection of PDoS attacks. Hence, intermediary sensor nodes block the dissipation of fabricated and/or re-played packets. Each packet destined from an end-point includes a completely new ``one-way hash chain'' number, thereby replaying is prevented this way.

\subsection{Summary of the Security Solutions}
Table~\ref{tab2} summarizes all the attacks against WSNs and IoT along with the proposed defense (detection, prevention or mitigation) solutions related to corresponding attacks. Understanding these attacks and their associated defense mechanisms will help researchers to provide a secure path in building public trust and acknowledgment while developing algorithms, systems, and concepts for IoT. 

Among those attacks, sinkhole and wormhole pose significant challenges to secure routing protocol design, and it is unlikely to devise effective countermeasures against these attacks that can be applied after the design of a protocol is completed. It is, therefore, crucial to design routing protocols in which these attacks are meaningless or ineffective. Geographic routing protocol \cite{gupta2015geographic} is one class of protocol that holds promise in this sense. 

An ultimate limitation of building a multi-hop routing topology around a fixed set of base stations is that those nodes within one or two hops of the base stations are particularly attractive for compromise. After a significant number of these nodes have been compromised, all the network is lost. This indicates that clustering protocols like LEACH \cite{heinzelman2000energy} and PCAC \cite{butun2015pcac}, where cluster-heads communicate directly with a base station, may ultimately yield the most secure solutions against node compromise and insider attacks.

%\newpage
\section{Challenges, Open Issues and Solutions on Cyber-Security of the IoT}\label{sec:iot}
\subsection{Technological Challenges}\label{sec:challenges}
Building up an IoT network is a challenge by itself and this provides us a path to understanding the overall challenges while devising security algorithms for IoT. As discussed in \cite{balte2015security}, challenges of IoT can be summarized as but are not limited to:  
\begin{itemize}
	\item \textbf{Heterogeneity:} IoT consists of a variety of devices belonging to various families such as gateways, switches, sensors, actuators, smart appliances, mobile systems, etc. These devices all run on different circuitry, use diverse protocols for communications, and employ distinct data processing algorithms. 
	\item \textbf{Scalability:} Addressing, naming, managing and servicing millions of devices is a unique challenge.
	\item \textbf{Communications:} Various technologies are used by IoT devices, such as wired or wireless communications e.g., Bluetooth, ZigBee, LPWAN.
	\item \textbf{Energy consumption:} This is one of the main challenging constraints of the IoT. Any kind of algorithm running on IoT devices needs to be designed with light-weight processing requirement.
	\item \textbf{Location privacy:} In the regular operation mode, things of the IoT should preserve their location privacy and when needed, they should provide this to network administrators.  
	\item \textbf{Self awareness:} Smart objects of the IoT should self-organize themselves autonomously in order to fulfill some pre-determined specific tasks in responding real-word environmental situations without too much human intervention.
	\item \textbf{Interoperability:} In order for heterogeneous IoT devices to communicate, collaborate and share data with each other, there should be a pre-determined and standardized data exchange format.  
\end{itemize}

\subsection{Open Issues}
The rapid development of IoT, Industrial IoT (IIoT) and Cyber-Physical Systems (CPS) has brought tremendous demand for smart-things (sensors, equipment, and devices, mostly referred as things) which are capable of sensing information from their surrounding, processing and transmitting it to far placed locations (mostly referred as data sinks) for further analysis and conclusions. Due to this extreme demand, cyber-security of these IoT enabled devices is somewhat disregarded \cite{li2017securing}. Therefore, both industrially and commercially used IoT devices are vulnerable to several classes of attacks and possess potential back-doors to the systems they have been attached to \cite{wurm2016security}. 

For instance, as discussed in \cite{sivaraman2015network}, smart home IoT-enabled appliances such as power switches, thermostat controls, smoke/fire alarms, ambient lighting systems, etc., bring security concerns to the attention of the public as they allow data sniffers to observe and conclude about the private activities of the house habitants. Apthorpe \textit{et al.} \cite{apthorpe2017spying} clearly has shown that an Internet Service Provider (ISP) or a network sniffer can easily infer inside activities of the house habitants and revoke their privacy even if they are employing encryption techniques to protect the content of data they are transmitting, by simply analyzing the Internet traffic pattern created by the smart home IoT devices.

Because of lacking rigid security precautions and bad user habits (sometimes the IoT implementers do not bother issuing a user and password for their devices, and sometimes they just continue with the default user name and password from the manufacturer), IoT devices are leveraged as a workforce of the botnets by ill mannered hackers. An example of this is \textit{QBot botnet}, which is also known as Bashlite, Gayfgt, Lizkebab and Torlus. This IoT botnet was discovered in 2014 with the source code published in 2015. Some variants of Qbot botnet reached over 100,000 infected devices, serving as the precursor to Mirai botnet \cite{qbot}.

\textit{Mirai malware} is another very good example to show the weaknesses of the things of IoT. \textit{Mirai malware} is devised against Linux OS based IoT devices and gains shell access of the devices to divert their operations towards the benefit of the \textit{Mirai botnet}. The Mirai botnet then uses this kind of captured zombie devices to perform further DDoS attacks against more advanced targets \cite{chaudhary2017privacy}. Most probably the users of those IoT devices won't even notice this unless inspected carefully. Therefore, in order to address this challenge, distributed and collaborative security solutions need to be optimized for IoT systems.

Recently revealed the \textit{Torii botnet}, is a cut above both the Mirai and QBot variants, according to researchers from Avast, as it possesses sophistication ``a level above anything we have seen before'' botnet attack has shown that security of IoT needs to be considered more seriously than ever \cite{torii}.    

Unfortunately, enslavement of thousands of IoT devices by botnets shows us that many IoT ecosystems do not even possess basic security elements. According to an investigation among commercial off the shelf IoT products \cite{bertino2017botnets}, the following are deduced: On average, 25 vulnerabilities are detected per device, 60\% had vulnerable interfaces and firmware, 70\% did not encrypt any communications at all, 80\% failed to request secure length-ed password for authentication. 

\subsection{Security Proposals}

\begin{table*}
	\caption{Promising cyber-security solutions for IoT}
	\centering
	\setlength{\tabcolsep}{3pt}
	\begin{tabular}{|p{165pt}|p{225pt}|}
		\hline
		\textbf{Security proposal}& 
		\textbf{Cyber-defense strategy}\\[0.2ex]
		%		\hline
		\hline
		ITS \cite{kumar2017securing} & Byzantine-resilient state machine approach for CPHS\\[0.2ex]
		PUFs \cite{xu2014security} & Hardware embedded security functions\\ [0.2ex]
		Privacy-preserving home automation network \cite{schurgot2015experiments} & Usage of encrypted overlay network over commercially available VPN\\[0.2ex] 
		Smart-home security\cite{dorri2017blockchain} & Block-chain technology in security and privacy of IoT\\[0.2ex] 
		6LowPSec \cite{glissa20186lowpsec}& End-to-end security solution that works on 6LoWPAN protocol\\[0.2ex] 
		CoAP protocol  \cite{shelby2011constrained} & Employment of DTLS which provide TLS equivalent security\\[0.2ex]  
		MQTT-Security \cite{neisse2014enforcement} &  Enforcement of security policy rules for IoT\\[0.2ex]
		SoftThings \cite{bhunia2017dynamic} &  Machine learning is used at the SDN controller\\[0.2ex]
		IoT security framework \cite{pacheco2016iot} & A threat model for IoT to be employed for smart cyber-infrastructures\\
		\hline
	\end{tabular}
	\label{tab3}
\end{table*}

IoT enabled Cyber-Physical Human Systems (CPHS) are one of the main components of the new era`s cyber cities. As mentioned in \cite{kumar2017securing}, the human is one of the key components of the CPHS. For the cyber-security of CPHS, besides IoT systems, human interactions with the CPHS needs to be considered. According to \cite{kumar2017securing}, $95\%$ of all security incidents happened due to human errors. Therefore, to defeat collaborative attacks against human errors, authors introduced an Intrusion Tolerant System (ITS) to support IoT enabled CPHS. The proposed ITS employs the Byzantine-resilient state machine approach, which combines replica diversity, voting, and cryptographic schemes to mask a number of compromised replicas of the nodes so that the CPHS can resume normal operation without distortion.

In \cite{schurgot2015experiments}, a privacy-preserving home automation network is proposed. An encrypted overlay network is created over commercially available Virtual Private Network (VPN) services to improve the privacy of the IoT users, especially the ones using the smart home systems.

In \cite{xu2014security}, authors proposed employment of hardware-based Physical Unclonable Functions (PUFs), to enhance and enable security-related operations to be handled at the sensor level in IoT. Usage of PUFs will help in increasing the security level of the IoT, by allowing low-level security implementations on the things and also by devising cryptography software to perform special tasks such as verification.  

To enhance the security of IoT, \cite{abera2016things} proposed remote attestation for trust establishment. According to authors, this is a non-trivial task because of the complexity of the design regarding the trust attestation schemes. These schemes, either hardware-based (e.g. PUFs) or software-based (e.g. control flow integrity checking), demand high power consumption along with extra economical cost, which is not suitable for vast implementation in the IoT. Some intermediate hybrid solutions (tailored in conjunction with the requirements and capabilities of the things) can be proposed in the near future, by blending both hardware- and software-based remote attestation schemes for trust establishment of IoT. These hybrid solutions need to be tested and verified for IoT, that they would able to work in scalable conditions and under the cyber-attacks such as DoS and against the condition of malicious verifiers.

Block-chain technology is a newly booming technology which was proposed for digital crypto-currencies. This technology was originally designed and invented by Bayer, Haber and Stornetta in 1992 \cite{bayer1993improving} and incorporated Merkle trees to provide the efficiency and reliability of the digital timestamps. As discussed in \cite{macdonald2018token}, F.A. Hayek published his classic book `Denationalization of Money' in 1976 \cite{von1976denationalisation} and argued that money is not different than other commodities and needs to be supplied by competition among private providers, not by the government. Crypto-currency has urged from that perspective, nowadays uses the block-chain technology and is a very hot topic. Besides, it attracted the public attention to the block-chain technology. Many researchers are working to bring and provide this technology for today's technological needs. Block-chain-based security algorithms provide a decentralized solution but involve significant energy, delay and computational overhead which is not suitable for resource constraint things of IoT. For instance, authors of \cite{dorri2017blockchain} have proposed usage of block-chain technology in security and privacy of IoT. In their proposal, authors employed high processing enabled miner devices, additionally attached to the home network, to provide needs and functionalities of the block-chain algorithms. However, proof-of-concept applications need to be developed and further analyzed in this manner.

A solution employing Software Defined Networking (SDN) is proposed by \cite{bhunia2017dynamic} to detect and mitigate dynamic attacks against IoT. In the proposed ``SoftThings'' framework, \textit{machine learning} is used at the SDN controller to monitor and learn the behavior pattern of IoT things over time. Anything out of the pre-determined behavior pattern is declared as an attack. The proposed scheme would be very effective on high processing capable IoT devices such as gateways and switches, but will not work on low-end IoT devices such as sensors and actuators.  

Pacheco and Hariri \cite{pacheco2016iot} proposed a threat model for IoT to be employed for smart cyber-infrastructures which helps IoT network administrators in identifying potential attacks against each layer. The proposed threat model consists of four layers: things, network, services, and applications.  

Providing security in IoT is challenging not only owing to the limited resources of the end-devices along with lossy communication links, but also due to the novel communications and networking technologies that are recently introduced such as RPL, 6LoWPAN, TSCH, MQTT, CoAP etc. Implications of using these technologies (one or many at the same time) under consideration of IoT network and device limitations need to be evaluated while taking security pre-cautions. In this manner, researchers are working in the field and Table~\ref{tab3} presents summary of promising cyber-security solutions that are proposed for IoT.

\section{Discussions and Final Remarks}\label{sec:discussion}
Section II provided all possible attack scenarios towards WSNs and IoT. Section III presented not only the detection techniques of attacks but also the prevention and mitigation techniques. Whereas, Section~\ref{sec:iot} specifically considered the cyber-security related defense strategies devised for IoT. All these sections revealed that devising a ``one-fits-all''  security solution to defend the WSNs and IoT against cyber-attacks is non-trivial. Therefore in this section, we provide partial possible candidate solutions from the literature that are targeting security of some specific features of the WSNs, especially IoT. These solutions can be categorized as follows; key distribution, trust management, data confidentiality, segmented attack discovery, location privacy, hierarchical intrusion detection, redundant data infusion.

\subsection{Key Distribution}
In order to enhance the security of WSNs against attacks, one should consider increasing the effectivity and flexibility of the key distribution scheme being used. When using encryption throughout WSNs; one of the critical issues is how to properly distribute the secret keys among sensor nodes when needed. Besides, revocation of the existing nodes and addition of new nodes are other challenges to be introduced while designing key management for WSNs. More interested readers would refer to \cite{chen2014survey, du2005pairwise, chan2003random, eschenauer2002key} for further detailed discussions regarding key distribution and key management in WSNs. 

\subsection{Trust Management}
Trust management is a very powerful concept to be used in detecting misbehaving sensor nodes in WSNs. After one of these nodes is detected by the trust evaluation, neighbors of this node can simply stop communicating with it. They can block the node by removing it from all routing tables, dropping all packets destined to/from it and stopping collaborating with it. By following this idea of using ``trust management'' to cope with attacks towards WSNs, many security methodologies and protocols are proposed in the literature. However, the proposed protocols introduced un-tolerable communication and/or computation overheads and unfortunately presented limited endurance against Sybil-attacks, DoS-attacks, and collusion-attacks. For instance, in the trust management scheme of \cite{han2014management}, the trust evaluation of sensor nodes depends on the previous behavioral pieces of evidence or the referrals from the neighboring nodes. However, predicting future trust of a sensor node depending on its historical course of trust is just ignored. To enhance accuracy and efficacy of trust evaluation, more trust metrics should be included, i.e. packet-loss, hop-count, radio transmission range, energy consumption rate, data link latency, path-quality, etc. 

As mentioned above and also in \cite{yu2012trust}, application of trust mechanisms seem to be the future of WSNs in order to secure the sensor network from attackers. Secure routing along with secure data aggregation techniques are being devised for WSNs, especially based upon the trust metrics of the nodes. An example would be the ``Trust-Aware Secure Routing Framework in Wireless Sensor Networks'' presented by Duan \textit{et al.} \cite{duan2014tsrf}, which employs the combination of trust and QoS metrics in terms of routing related metrics in an effort to provide an optimized and qualified routing algorithm for WSNs.

In \cite{xiang2012research}, Xiang \textit{et al.} presented ``Addition Encouragement, Multiplication Punishment (AEMP)'' trust model in which each node in the network can calculate communication trust values of all its neighbors. An enhanced version of the AEMP trust model can be devised for WSNs in order to fight against the increasing number of attack scenarios. 

Secure and efficient data transmission as shown in \cite{lu2014secure} would be another solution avenue to prevent attacks against WSNs. Encryption prevents eavesdropping and impersonation attacks but it brings an extra computational load to the sensor nodes, which in most cases have strict power constraints. 

ActiveTrust of \cite{liu2016activetrust} is also another promising security solution for WSNs, in which trusted routing is achieved through an active detection route protocol. In ActiveTrust, basically the location and behavior of an attacker, along with the node related trust, could be gathered and utilized for avoiding blackholes while in process of real-data route construction. This provides ``blackhole'' free paths for the real-data traffic and improves network lifetime as well as network QoS drastically.

\subsection{Data Aggregation}
In WSNs, owing to the limited energy sources and computation power, data aggregation among sensor nodes is achieved at the aggregating-node and generally, this is executed by using simple methodologies like as data-averaging. Nonetheless, data-aggregation is not robust against node-compromisation attacks. Hence not only sensor nodes are generally deployed in unattended environments, but they also lack tamper-proof hardware; they are very vulnerable to those mentioned attacks. Therefore, the inclusion of trustworthiness to data management and reputing mechanisms of nodes has prime importance for sensor networks as well as IoT. Secure data aggregation techniques such as Iterative Filtering (this algorithm is an appealing alternative for sensor networks hence it solves both problems 1) data aggregation, 2) assessment of data-trustworthiness; by employing a unique iterative method \cite{rezvani2015secure}), would be used in WSNs and IoT in order to cope with the vulnerabilities mentioned above.

Therefore, a unique way of providing security to WSNs and IoT would be ``secure data aggregation''. In the specific applications where average sensed data is of the interest, for example, the average temperature of a crop field, this method may be used efficiently. Data can be averaged at some ``aggregation points'' and these aggregation points might communicate each other in a more secure way by implementing ways of Secret Key Cryptography (SKC). This way, more attention can be paid to this specific data gathering points and hence more secure data dissemination can be achieved. 

\subsection{Hierarchical Security Solution}
Another promising technique to cope with attacks against WSNs would be hierarchical security solutions:

Butun \textit{et al.} \cite{butun2015intrusion}'s hierarchical IDS provides two different paths of detecting intruders through watchdog and majority voting mechanisms. Multi-level cluster heads and subordinate nodes are being watched and recorded separately, and at a specific threshold level, intruders have revoked from the network accordingly. 

In Wu \textit{et al.} \cite{wu2016hierarchical}'s hierarchical security framework, authors employed dynamic adaptive chance discovery mechanism to detect unknown attacks. With this unified framework, low-level attack-detection is executed in sensor nodes with simple rules, and high-level attack-detection is executed in sinks and at the base-station with complex rules. Besides, software-defined networking and network function virtualization technologies are used to perform attack mitigation when any type of attack is detected. 

\subsection{Traffic Shaping}
Independent Link Padding (ILP) is a network traffic shaping methodology proposed by Apthorpe \textit{et al.} \cite{apthorpe2017spying} in an effort to protect smart home IoT systems from network sniffing (eavesdropping) attacks. The proposed ILP scheme is promising in protecting privacies of the house habitants from unintended eyes. ILP works in a sense to shape the traffic without violating pre-determined data rate and schedule of the regular network traffic so that it does not cause disturbance or loss for the data communication activities of the smart home IoT devices. In essence, the result of the ILP procedure ends up with a constant flow of information in which regular and meaningful data packets are padded with redundant data so that no useful information can be interpolated from the data traffic. By this way, the privacy of the smart home users is assured against passive information gathering (sniffing or eavesdropping) attack. ILP methodology improves the privacy of the IoT users, however, it causes bandwidth of the network to be wasted by introducing extra load and causing unnecessary traffic.  

\subsection{Patch management} Patch management possesses a very significant role in industrial networks and IIoT, in which timely patching firmware of the critical devices has prime importance, such as the ones in SCADA networks, that are deployed over critical infrastructures and factories \cite{butun2012networking}. As mentioned in \cite{stallings2018computer}, patching vulnerabilities of the IoT is also very important as in the case of many computer networks, however, owing to their low-cost components, i.e. things of IoT consists of very cheap embedded devices most of which do not have upgradeable firmware, this is quite impossible. 

\subsection{Automated fingerprinting}
In order to prevent node replication attacks and variants towards IoT, an automated fingerprinting technique can be used. In this technique, several messages emanating from legitimate devices can be characterized, such as periodic status update messages, firmware update messages, patch update messages, device initiation messages, etc. Any behavior, out of this fingerprint can be evaluated as a suspicious act or an attack.

\section{Conclusion}\label{sec:conclusion}
The IoT is a large-scale complex architectural design consisting of a variety of heterogeneous devices, therefore scalability, transparency, and reliability are most prominent issues to be solved. Security-related initiatives need to consider these issues in the first place. Besides,  not only should a higher architectural security design be conceptualized, but low-level security also needs to be addressed. This can be achieved by designing lightweight security protocols and cryptography algorithms that are tailored according to the specific needs of the resource-constrained devices of the IoT.

From the discussions made throughout this paper, it can be deduced that the heterogeneity of the devices in IoT ecosystem along with its scalability causes several implications, in terms of security. Although some of the new vulnerabilities can be discovered on time, related security patches cannot be installed to the end devices in a timely manner due to the mentioned IoT network implications above. Therefore IDS techniques become more important for IoT systems, as some of them are even efficient against zero-day-attacks. If the necessary IDS techniques require high processing power, gateway devices can be employed for this purpose. 

As a conclusion, security must be a key component when designing protocols for WSNs as well as IoT. Without a proper assessment of possible threats and inclusion of related preventive measures, these networks will be vulnerable to attacks. Future researchers working on WSNs and IoT need to consider security to a higher extent while designing their routing, key distribution, trust management, and finally data aggregation schemes over MAC, networking, transport, and application layers.This paper has compiled all known types of security attacks towards WSNs in IoT context together with a description and evaluation of the defending strategies against each type of attack. Authors hope that this article will shed light on the researchers working in the field of WSNs and IoT, by leading them to produce more robust and secure network solutions.

\appendices
\section{Abbreviations and Acronyms}
List of abbreviations are listed in Table~\ref{tab4}.

\begin{table}
	\caption{List of abbreviations.}
	\centering
	\setlength{\tabcolsep}{3pt}
	\begin{tabular}{|p{43pt}|p{189pt}|}
		\hline
		\textbf{Abbreviation}& 
		\textbf{Explanation}\\ [0.2ex]
		\hline
		ARP & Address Resolution Protocol\\
		CIA & Confidentiality, Integrity, Availability\\
		CoAP & Constrained Application Protocol\\
		CoAPs & CoAP in secure mode (by using DTLS)\\
		CPS & Cyber Physical System\\
		CPHS & Cyber Physical Human System\\
		CTS & Clear to Send \\ 
		DODAG & Destination Oriented Directed Acyclic Graph\\
		DIO & DODAG Information Object\\
		DoS & Denial of Service\\ 
		DTLS & Datagram Transport Layer Security\\
		ID & Identity\\
		IDS & Intrusion Detection System\\
		ILP & Independent Link Padding\\
		IoT & Internet of Things\\
		IPS & Intrusion Prevention System\\
		ISP & Internet Service Provider\\
		MAC & Medium Access Control\\
		MANET & Mobile Ad-Hoc Network\\
		MEMS & Micro Electro Mechanical Systems\\
		MQTT & Message Queue Telemetry Transport\\
		OSI &  Open Systems Interconnection\\
		PKC & Public Key Cryptography\\
		PUF & Physical Unclonable Function\\
		PZ & Promiscuous Zone\\
		RPL & Routing Protocol for Low-power and lossy networks\\
		RREQ & Route Request\\
		RREP & Route Reply\\
		RTS & Request to Send\\
		SDN & Software Defined Networking\\ 
		SKC & Secret Key Cryptography\\
		TLS & Transport Layer Security\\
		TSCH & Time Slotted Channel Hopping\\
		VPN & Virtual Private Network\\
		WSN & Wireless Sensor Network\\		  
		6LowPAN & Internet protocol devised for the IoT for the usage with IPv6\\
		6TiSCH&IPv6 over the TSCH mode\\
		\hline
	\end{tabular}
	\label{tab4}
\end{table}

\ifCLASSOPTIONcaptionsoff
  \newpage
\fi



%\bibliographystyle{IEEEtran}

%\begin{thebibliography}{1}
%\end{thebibliography}

\bibliographystyle{IEEEtran} %butun
\bibliography{IEEEabrv,butun} %butun


\newpage

\begin{IEEEbiography}%[{\includegraphics[width=1in,height=1.25in,clip,keepaspectratio]{butun.jpg}}]
{Ismail Butun} 
(M'09) received his B.Sc. and M.Sc. degrees in Electrical and Electronics Engineering from Hacettepe University, his M.Sc. and Ph.D. degrees in Electrical Engineering from the University of South Florida. He worked as an Assistant Professor in years between 2015 and 2016. Since 2017, he has been working as a post-doctoral fellow for various universities (University of Delaware, Mid Sweden University, Chalmers University of Technology). He has more than 30 publications in peer-reviewed scientific international journals and conference proceedings, along with an H-index of 11. He is a well recognized academic reviewer by IEEE, ACM, and Springer, who served for 32 various scientific journals and conferences in the review process of more than 80 articles. He is an editor of Springer Nature. His research interests include but not limited to; computer networks, wireless communications, WSNs, IoT, cyber-physical systems, cryptography, network security, and intrusion detection.

\end{IEEEbiography}

\begin{IEEEbiography}%[{\includegraphics[width=1in,height=1.25in,clip,keepaspectratio]{oster.jpg}}]
{Patrik \"{O}sterberg} (M'05) received his M.Sc. degree in Electrical Engineering from Mid Sweden University, Sundsvall, Sweden, in 2000, the degree of Licentiate of Technology in Teleinformatics from the Royal Institute of Technology, Stockholm, Sweden, in 2005, and the Ph.D. degree in Computer and System Science from Mid Sweden University in 2008. During 2007, he worked as a development engineer at Acreo AB in Hudiksvall, Sweden, and from 2008 to 2010, he was employed as researcher at Interactive TV Arena KB in G$\ddot{a}$vle, Sweden. Since 2008, he is an Assistant Professor at Mid Sweden University and from 2013, he is also the head of the Department of Information System and Technology.
\end{IEEEbiography}

\begin{IEEEbiography}%[{\includegraphics[width=1in,height=1.25in,clip,keepaspectratio]{hsong}}]
{Houbing Song}
(M'12--SM'14) received the Ph.D. degree in electrical engineering from the University of Virginia, Charlottesville, VA, in August 2012, and the M.S. degree in civil engineering from the University of Texas, El Paso, TX, in December 2006.
In August 2017, he joined the Department of Electrical, Computer, Software, and Systems Engineering, Embry-Riddle Aeronautical University, Daytona Beach, FL, where he is currently an Assistant Professor and the Director of the Security and Optimization for Networked Globe Laboratory (SONG Lab, www.SONGLab.us). He served on the faculty of West Virginia University from August 2012 to August 2017. In 2007 he was an Engineering Research Associate with the Texas A\&M Transportation Institute. He serves as an Associate Technical Editor for IEEE Communications Magazine. He is the editor of four books, including Smart Cities: Foundations, Principles and Applications, Hoboken, NJ: Wiley, 2017, Security and Privacy in Cyber-Physical Systems: Foundations, Principles and Applications, Chichester, UK: Wiley-IEEE Press, 2017, Cyber-Physical Systems: Foundations, Principles and Applications, Boston, MA: Academic Press, 2016, and Industrial Internet of Things: Cybermanufacturing Systems, Cham, Switzerland: Springer, 2016.  He is the author of more than 100 articles. His research interests include cyber-physical systems, cybersecurity and privacy, internet of things, edge computing, big data analytics, unmanned aircraft systems, connected vehicle, smart and connected health, and wireless communications and networking.
Dr. Song is a senior member of ACM. Dr. Song was a recipient of the prestigious Air Force Research Laboratory's Information Directorate (AFRL/RI) Visiting Faculty Research Fellowship in 2018, and the very first recipient of the Golden Bear Scholar Award, the highest campus-wide recognition for research excellence at West Virginia University Institute of Technology (WVU Tech), in 2016. 
\end{IEEEbiography}

%% Balancing the columns of the last page
%\balance

%% insert where needed to balance the two columns on the last page with
%%\newpage


%\vfill


\end{document}